\documentclass[twocolumn, trackchanges]{aastex701}

\shorttitle{Towards Automated Scientific Discovery}
\shortauthors{Sun et al.}
\submitjournal{AAS Journals}

\usepackage{mdframed} 
\usepackage{array}   

\newmdenv[
    linecolor=black,     
    linewidth=2pt,      
    backgroundcolor=gray!10, 
    roundcorner=10pt,    
    innertopmargin=10pt, 
    innerbottommargin=10pt, 
    innerleftmargin=10pt, 
    innerrightmargin=10pt 
]{algobox}

\newcommand{\kw}[1]{\textbf{#1}}      
\newcommand{\com}[1]{$\quad$$\quad$ $\triangleright\,\,${#1}}   

\usepackage{amsmath,amssymb,amsfonts}
\usepackage{graphicx}
\usepackage{pifont}
\usepackage{wasysym}
\usepackage{textcomp}
\usepackage{xcolor}
\usepackage{verbatim}
\usepackage{textcomp}
\usepackage{xcolor}
\usepackage{subfig}
\usepackage{verbatim}
\usepackage[most]{tcolorbox}
\usepackage[utf8]{inputenc}
\usepackage{listings}
\usepackage{float}
\usepackage{natbib}

\newtcolorbox{questionbox}[1][]{
  colback=#1,
  colframe=black,
  boxrule=1pt,
  arc=5pt,
  outer arc=5pt,
  enhanced,
  breakable
}

\definecolor{boxcolor1}{RGB}{230,240,255} 
\definecolor{boxcolor2}{RGB}{255,240,230} 
\definecolor{boxcolor3}{RGB}{230,255,230} 
\definecolor{boxcolor4}{RGB}{255,230,255} 
\definecolor{boxcolor5}{RGB}{255,255,230} 
\definecolor{boxcolor6}{RGB}{230,255,255} 

\begin{document}

\title{Mephisto: Self-Improving Large Language Model-Based Agents for Automated Interpretation\\ of Multi-band Galaxy Observations}

\author[orcid=0000-0002-8246-7792]{Zechang Sun}
\affiliation{Department of Astronomy, Tsinghua University, Beijing, China}
\email[show]{szc22@mails.tsinghua.edu.cn, ting.74@osu.edu, \\ shuang@mail.tsinghua.edu.cn}

\author[orcid=0000-0001-5082-9536]{Yuan-Sen Ting} 
\affiliation{The Ohio State University, Columbus, OH 43210, USA} 
\affiliation{Center for Cosmology and AstroParticle Physics (CCAPP), The Ohio State University, Columbus, OH 43210, USA}
\email{ting.74@osu.edu}

\author[orcid=0000-0002-6595-5145]{Yaobo Liang}
\affiliation{Microsoft Research Asia, Beijing, China}
\email{yalia@microsoft.com}

\author[orcid=0000-0002-3387-4674]{Nan Duan}
\affiliation{Microsoft Research Asia, Beijing, China}
\email{nanduan.nlp@outlook.com}

\author[orcid=0000-0003-1385-7591]{Song Huang}
\affiliation{Department of Astronomy, Tsinghua University, Beijing, China}
\email{shuang@mail.tsinghua.edu.cn}

\author[orcid=0000-0003-1385-7591]{Zheng Cai}
\affiliation{Department of Astronomy, Tsinghua University, Beijing, China}
\email{zcai@mail.tsinghua.edu.cn}

\correspondingauthor{Zechang Sun,Yuan-Sen Ting,Song Huang}

\begin{abstract}

Astronomical research has long relied on human expertise to interpret complex data and formulate scientific hypotheses. In this study, we introduce Mephisto—a multi-agent collaboration framework powered by large language models (LLMs) that emulates human-like reasoning for analyzing multi-band galaxy observations. Mephisto interfaces with the CIGALE codebase (a library of spectral energy distribution, SED, models) to iteratively refine physical models against observational data. It conducts deliberate reasoning via tree search, accumulates knowledge through self-play, and dynamically updates its knowledge base. Validated across diverse galaxy populations—including the James Webb Space Telescope’s recently discovered “Little Red Dot” galaxies-We show that Mephisto demonstrates ability in inferring the physical properties of galaxies from multi-band photometry, positioning it as a research copilot for astronomers. Unlike prior black-box machine learning approaches in astronomy, Mephisto offers a transparent, human-aligned reasoning process that integrates with existing research practices. This work underscores the possibility of LLM-driven agent-based research for astronomy, establishes a foundation for fully automated, end-to-end artificial intelligence (AI)-powered scientific workflows, and opens up new avenues for AI-augmented discoveries in astronomy.

\end{abstract}



\section{Introduction}\label{sec:intro}




The advent of deep learning tools, combined with the vast amount of data routinely collected in astronomical surveys—from hundreds of millions of spectra \cite[e.g., ][]{SDSS2000,GAIA2023,DESIEDR2024} to tens of billions of images \cite[e.g., ][]{LSST2019,EUCLID2022,HSC2022}—has advanced astronomical research into new territory. However, most AI applications in astronomical research have thus far focused on optimizing individual downstream tasks: building classifiers and brokers to streamline surveys \cite[e.g., ][]{BEN2022,STRONGLENSING2022,BRANT2023}, emulating computationally expensive hydrodynamical simulations \cite[e.g., ][]{CHARDIN2019, DAI2021, CABAYOL2023}, and advancing statistical inference using generative models as posterior and likelihood surrogates \cite[e.g., ][]{XIAOSHENG2022,SUN2023a,BINGJIE2023,JIAXUAN2024}. 

Yet despite these advances, challenges in interpretability, limited generalization to unseen data, and the focus on isolated tasks have constrained AI's broader impact in astronomy \cite[e.g., ][]{BARON2019,FLUKE2020,HOGG2024}. The recent emergence of large language models offers new avenues for scientific research \cite[e.g., ][]{YOSH2023,BIRHANE2023,LEI2024,JAB2024,MICROSOFT2023}. These models have demonstrated  capabilities in scientific reasoning across diverse fields \cite[e.g., ][]{AZERBAYEV2023,SINGHAL2023,JABLONKA2024}, leading to the development of LLM-based agents that can interact with their environment, make decisions, and perform actions \cite[e.g.,][]{AUTOGEN2023,WANGLEI2023,TAICHENGGUO2024}. While such agents have shown success in applications ranging from drug discovery \cite[e.g.,][]{CHEMCROW2023,GAOSHANGHUA2024,JIASHUYI2024} to software development \cite[e.g.,][]{VALLECILLOS2024,ARORA2024,LIUJUNWEI2024}, their potential for addressing the complex reasoning challenges in astronomical research remains largely unexplored.

The potential of LLM-based agents is relevant for astronomy, where the scientific process faces distinct methodological challenges compared to fields like biochemistry \cite[e.g., ][]{JUMPER2021,TUNYASUVU2021,MADANI2023,NOTIN2024} or materials research \cite[][]{PYZERKNAPP2022,MERCHANT2023,MANIC2023,TRAN2024}. While these disciplines often leverage controlled laboratory experiments, astronomy primarily relies on observational data to construct and validate causal models that explain celestial phenomena \cite[e.g., ][]{BIRNEY2006,PEARL2009,BURNS2023}. This observational approach, centered on proposing and refining theoretical frameworks through careful analysis of astronomical data, has been instrumental in developing our understanding of physics \cite[e.g.,][]{FUKU2004,BERTONE2005,FRIEMAN2008}.

Galaxies exemplify both the importance and complexity of this modeling approach. As building blocks of the Universe \cite[e.g., ][]{MO2010,VOGEL2020}, they serve as cosmic laboratories for investigating important questions about dark matter, cosmic expansion, black hole evolution, and star formation \cite[e.g., ][]{VALL2024,DESIY12024,MOUNTRICHAS2023,DOBBS2014}. Understanding their properties requires sophisticated analysis of spectral energy distributions (SEDs) \cite[e.g., ][]{CONROY2013}, combining multiple physical components: star formation history \cite[e.g.,][]{JOHN2021}, stellar populations \cite[e.g., ][]{BRUZUAL2003}, dust effects \cite[e.g.,][]{CALZETTI2000}, nebular emission \cite[e.g.,][]{FERLAND2013}, and active galactic nuclei contributions \cite[e.g.,][]{FRITZ2006}.

While high-quality spectroscopic observations would provide detailed information about distant galaxies, obtaining such data is resource-intensive. Instead, astronomers often rely on multi-band photometry—measurements of galaxy light through different filters—to construct coarse, low-resolution spectra \cite[e.g., ][]{SEDFIT2012,CONROY2013,CHEV2016,CIGALE2019}. This approach introduces technical challenges beyond mere mathematical optimization. Photometric measurements contain both random noise and systematic errors from instrumental effects \cite[][]{ZINE2022}, while galaxies themselves exhibit diversity in their physical properties and evolutionary histories. The interpretation of these sparse observations often creates competing solutions: multiple combinations of physical parameters and assumptions can match the data equally well while suggesting radically different interpretations.

For example, identifying high-value candidates—such as high-redshift galaxies and Population III galaxy candidates—from billions of observational datasets demands both a deep understanding of the underlying physical process and extensive expertise in processing observational data \cite[e.g.,][]{OESCH2014,FUJIMOTO2025}. There is no universal protocol for interpreting galaxy properties via spectral energy distribution (SED) modeling, nor is this task merely a matter of optimizing over a broad superset of parameters. Instead, distinct scientific questions and physical scenarios have driven the development of specialized models and fitting routines. Success hinges on an astronomer’s ability to navigate multiple modeling frameworks: directing observational data to the most appropriate models, evaluating competing physical scenarios, and making well-informed judgments about which underlying assumptions to revise.

This human-centered approach does not scale well to the volume of data in the era of modern surveys. Even observations from modern facilities like the James Webb Space Telescope—whose high-redshift galaxy observations continue to test our understanding of early galaxy formation \cite[e.g.,][]{LABBE2023,XIAO2023,GREENE2024}—see only a fraction of their data receive detailed analysis. The majority of sources undergo only basic evaluation through heuristic approaches for tri-aging, potentially missing galaxies that could reveal unknown physical processes.

To address these challenges, we introduce Mephisto, a large language model-based agent framework designed to emulate human reasoning in astronomical research. The system iteratively explores hypothesis spaces by proposing new models, interacting with simulation tools for validation, and reflecting on limitations to refine physical assumptions, much like human astronomers. Mephisto develops analytical intuition for potential solutions without relying on extensive pre-existing training data, making it particularly effective for addressing frontier questions. Its natural language interface ensures full transparency in its reasoning, allowing researchers to understand, validate, and guide its decision-making process. Additionally, its active explorations, guided by analytical intuition, equip it to handle frontier scientific inquiries without dependence on extensive pre-existing training data. 

We detail Mephisto’s complete framework in Section~\ref{sec:method}, including its core components: state representation, tree-based reasoning workflow, dual memory systems, and integration with the CIGALE SED modeling codebase. Section~\ref{sec:result} presents our key experimental findings: Mephisto’s SED fitting performance across diverse COSMOS2020 galaxies, its analysis of  "Little Red Dots", ablation studies quantifying the impact of memory mechanisms, and cost-efficiency comparisons between different LLMs to assess scalability. In Section~\ref{sec:discussion}, we identify the framework’s current limitations and outline potential future enhancements for agents designed for automated scientific discovery. Finally, we conclude in Section~\ref{sec:con}.

\section{Mephisto -- A Multi-Agent Framework for Interpreting Multi-band Galaxy Observations}\label{sec:method}

\begin{figure*}
    \centering
    \includegraphics[width=0.95\linewidth]{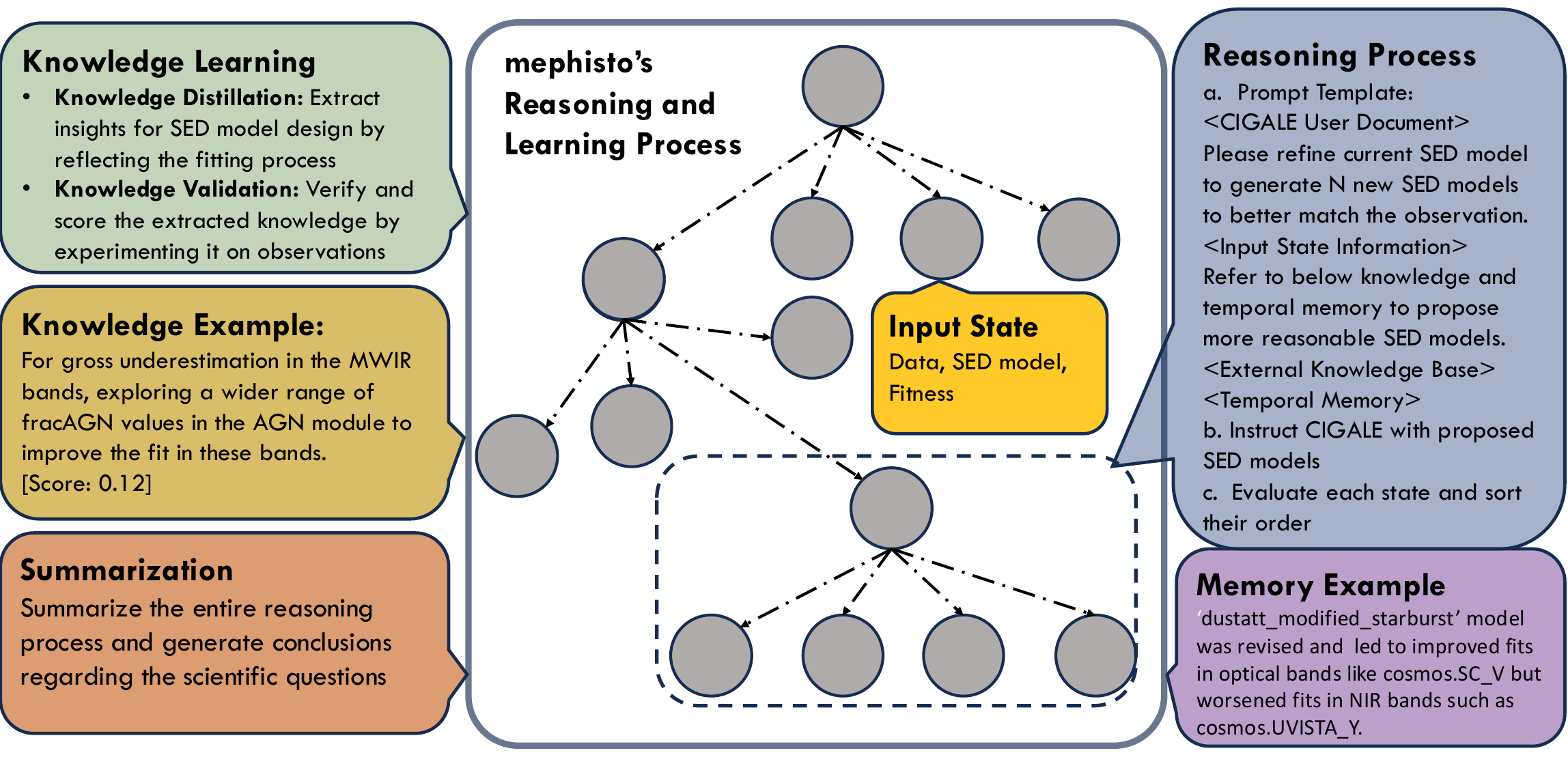}
    \caption{Overview of Mephisto's reasoning and learning process for interpreting multi-band galaxy observations. The central diagram shows Mephisto's tree-based exploration of different SED models, starting from an initial state (Input State) that includes observational data, the current SED model, and fitness metrics. The Reasoning Process (right) follows a structured approach: (a) using prompt templates that incorporate CIGALE documentation to propose model refinements, (b) instructing CIGALE with the proposed models, and (c) evaluating and prioritizing different states. This process is enhanced by two learning components: Knowledge Learning (top left) continuously extracts and validates insights from fitting experiences through knowledge distillation and validation (shown with a concrete example for AGN parameter adjustment), while Temporal Memory (bottom right) tracks the impact of previous model modifications. The entire analysis culminates in a scientific report through the Summarization component (bottom left), synthesizing the complete reasoning chain and conclusions.}
    \label{fig:schema}
\end{figure*}

\begin{figure*}[ht!]
    \centering
    \includegraphics[width=0.95\linewidth]{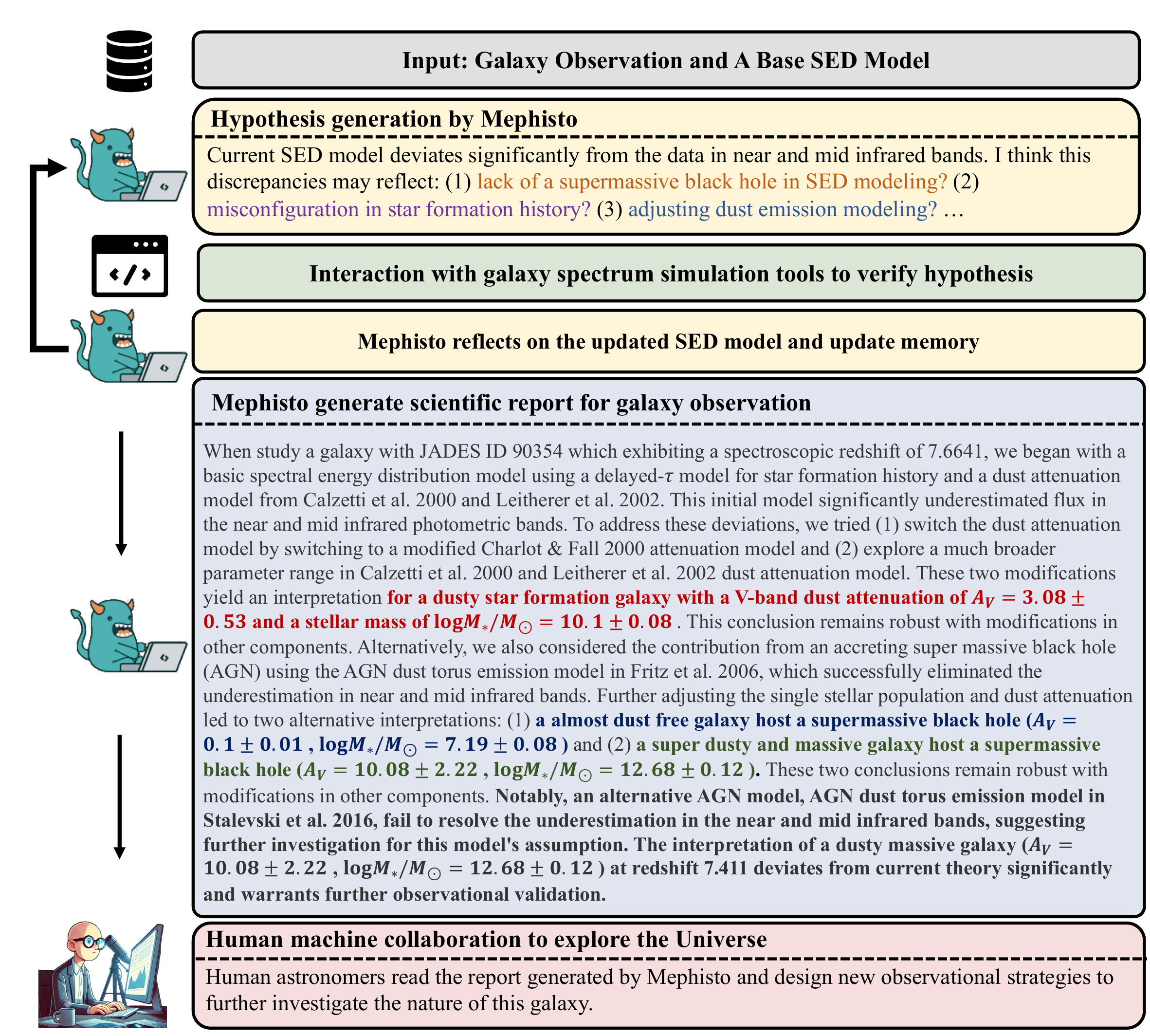}
    \caption{Mephisto's analysis of a special class of outliers known as Little Red Dots, recently discovered by the James Webb Space Telescope (observational data: JADES ID 90354 from \cite{PABLO2024}). Since these sources were only discovered in 2024, likely beyond most LLM knowledge cutoff dates, and their physical nature remains debated, they serve as a valuable test case demonstrating how Mephisto can navigate and identify solutions for objects through reasoning about physical processes not well-represented in LLM training data. Given only multi-band photometry and a base SED model, Mephisto iteratively explored and refined the physical model, developing explanations that more closely align with the observed properties of the galaxy. Throughout this exploration, Mephisto not only enriched the space of potential hypotheses for the current observations, but also validated the robustness of scientific conclusions across different model selections. Rather than merely providing parameter estimates, Mephisto produces a scientific report distilled from its reasoning process, encoding information about possible interpretations of the observations and the consistency between different models.}
    \label{fig:reasoning}
\end{figure*}

Pre-trained on trillions of tokens with billions of parameters, Large Language Models (LLMs) have demonstrated generalization capabilities in natural language processing tasks, ranging from complex question answering to coherent multi-step reasoning. This process of learning from vast and diverse data not only enables them to acquire a rich, if general, knowledge base of the world but also the ability to manipulate that knowledge through structured reasoning chains. It is this capacity for logical inference and planning that makes LLMs effective as the core of autonomous agents—systems that can perceive their environment, reason about goals, and execute actions to achieve them. We therefore posit that an LLM agent, when equipped with domain-specific tools and guided by appropriate reasoning structures, can emulate the intricate, iterative process of scientific inquiry required for interpreting galaxy observations.

This potential is precisely what we harness in Mephisto, which is designed to emulate the scientific reasoning process for galaxy observations that would traditionally require human expertise. The interpretation of galaxy SEDs requires intricate reasoning that combines theoretical knowledge, observational expertise, and iterative model refinement. Mephisto is designed to emulate this scientific reasoning process, particularly for galaxy observations that would traditionally require human expertise. Since different galaxies can be dominated by vastly different physical processes—from intense star formation to active galactic nuclei—SED fitting first requires determining which physical components to include and their underlying assumptions. This choice determines the appropriate codebase and fitting approach. Unlike traditional optimization problems where one searches over a fixed parameter space, SED fitting involves exploring different model combinations where the parameter spaces themselves can vary with different assumptions. Mephisto approaches this challenge as a scientific investigation, developing analysis strategies through agentic reflection, rather than attempting brute force optimization across all physically plausible options. This approach mirrors how astronomers systematically explore and refine physical models to explain observational data.

Mephisto's primary function is to analyze multi-band photometric observations, typically comprising 20 to 40 data points across various wavelengths, and to generate scientific reports detailing the possible physical scenarios of the input galaxies. The framework interfaces with CIGALE\footnote{\href{https://cigale.lam.fr}{https://cigale.lam.fr}} \cite[][]{CIGALE2019}, a widely adopted SED modeling software that provides a library of physical models. The SED model incorporates these physical components available in CIGALE: star formation histories \cite[][]{CIGALE2019}, stellar populations \cite[e.g., ][]{BRUZUAL2003,MARASTON2005}, dust attenuation and emission \cite[e.g., ][]{LEITHERER2002,CALZETTI2000,DALE2014}, nebular emissions \cite[e.g., ][]{FERLAND1998,FERLAND2013}, and AGN contributions \cite[e.g., ][]{FRITZ2006,CAMPS2015}. CIGALE's unified configuration system allows Mephisto to systematically modify physical assumptions while maintaining consistent interfaces and outputs.

Starting with minimal assumptions, Mephisto attempts to analyze the discrepancies between this initial model (detailed at Appendix~\ref{appendix:base_model}) and observation. Through an iterative process, as shown in Figure~\ref{fig:schema}, it builds increasingly sophisticated models by examining these discrepancies and systematically incorporating or modifying various physical components as needed.

To formalize this iterative process, Mephisto operates on a state representation $s(d, m, r)$, which combines three key components: observational data $d$, the current SED model $m$, and fitting results $r$. The observational data consists of the galaxy's redshift and photometric measurements, each characterized by effective wavelength, bandwidth, flux value, and signal-to-noise ratio. While human astronomers heavily rely on visual interpretation of SED plots, current vision-enabled large language models remain limited in their ability to understand scientific plots quantitatively \cite[e.g., ][]{EBRAHIMIKAHOU2017,MASRY2022,YUE2023}. Therefore, we represent the data as tuples that integrate numerical values with annotated tags, prioritizing analysis with LLM agents over replicating human visual intuition (See Appendix~\ref{appendix:input_prompt} for detailed implementations for input state). 

For each state, Mephisto analyzes discrepancies between model predictions and observations, generating $\mathrm{N}_\mathrm{b}$ hypotheses for model refinement (we chose $\mathrm{N}_\mathrm{b}=4$ in this study) (See Appendix~\ref{appendix:reasoning_prompt} for reasoning prompt). Building on established approaches to LLM reasoning \cite[][]{YAO2023,GOT2023,AOT2023,CHEN2022}, we adopt a tree structure to facilitate deliberate model exploration. Starting from basic assumptions—similar to how astronomers approach unknown sources \cite[][]{WAL2011,SEDFIT2012,PACI2023}—Mephisto systematically modifies physical components as needed. We implement both depth-first search (DFS) and breadth-first search (BFS) strategies: DFS pursues promising directions by expanding the best-performing nodes, while BFS ensures broader exploration by investigating alternative paths \cite[][]{YAO2023} (see Appendix~\ref{appendix:code}).

The fitting results provide an evaluation of model performance through a combination of qualitative and quantitative metrics that address the inherent challenges of photometric data analysis. Each photometric band receives a qualitative assessment (categorized as \texttt{good}, \texttt{overestimate}, or \texttt{underestimate}) alongside traditional metrics like parameter estimates, residuals, and computational costs. This qualitative approach serves multiple important purposes. First, it mitigates the known tendency of large language models to hallucinate when handling numerical tasks \cite[][]{HENDRYCKS2021,FRIEDERSIMON2023,AHN2024} by transforming quantitative discrepancies into discrete categories. Second, it accommodates the reality of astronomical data, where controlled experiments are not possible and observations may contain anomalous data points (e.g., from cosmic rays) that might not be fully captured in uncertainty assessments \cite[][]{ZINE2022}. Finally, it harnesses insights from how astronomers typically communicate and reason about SED fitting in the research literature, enabling Mephisto to better leverage domain knowledge by recognizing patterns—for example, understanding that poor fits in ultraviolet bands might suggest the need to adjust starburst components.

Rather than prescribing fixed evaluation criteria, we provide Mephisto with multiple metrics through its prompt template, allowing it to prioritize these factors based on context and astronomical knowledge. In practice, we encourage Mephisto to prioritize robustness by first considering the number of bands where model predictions fall within $1\,\sigma$ of observational uncertainties, using $\chi^2$ as a secondary criterion when models achieve similar band-wise performance. This approach reflects established astronomical practice where reliable fits across multiple bands often provide more physical insight than optimizing a single global metric, as different wavelength regions probe distinct physical processes within galaxies.

\begin{figure*}[htbp]
    \centering
    \gridline{
    \fig{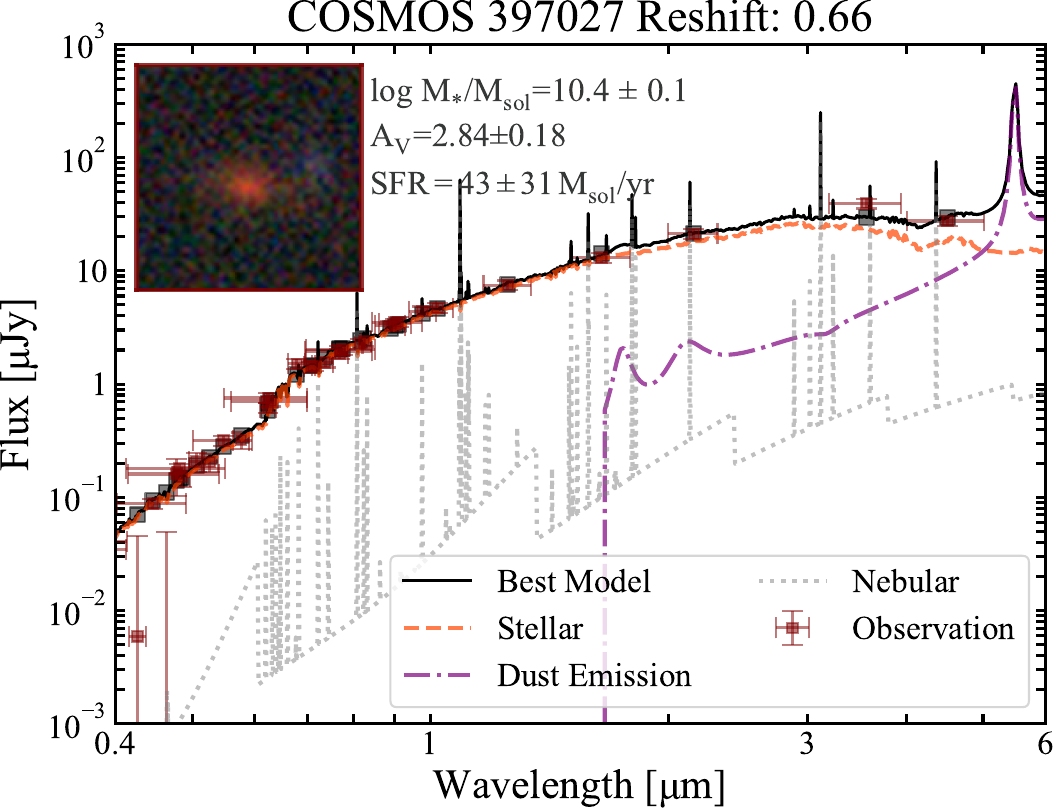}{0.3\textwidth}{A Dusty Star-forming Galaxy \label{fig:cosmos_demo_a}}
    \fig{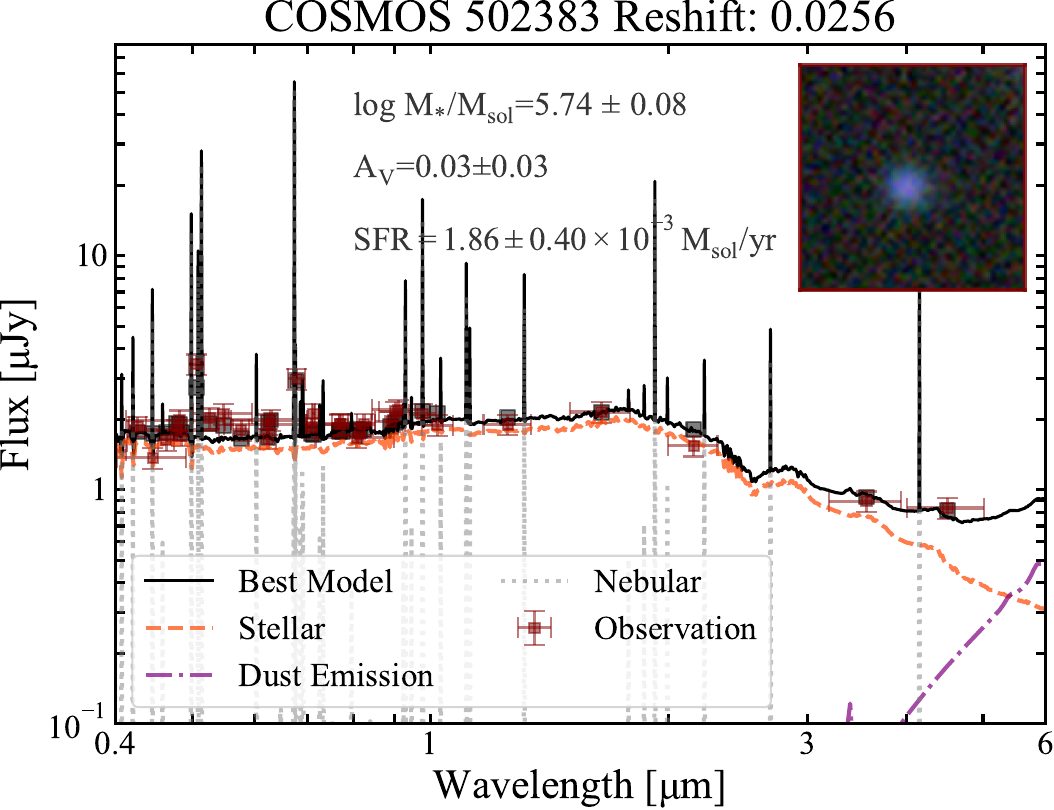}{0.3\textwidth}{A Dwarf Galaxy\label{fig:cosmos_demo_b}}
    \fig{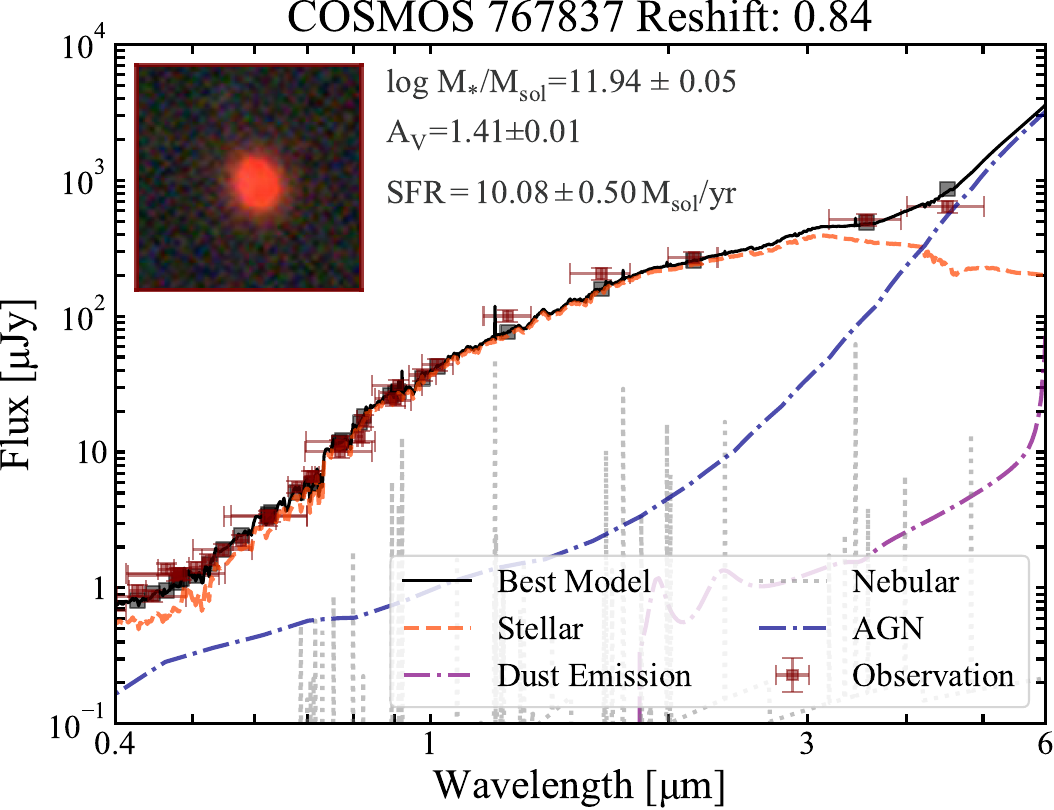}{0.3\textwidth}{A Massive Galaxy with Possible AGN Contribution\label{fig:cosmos_demo_c}}
    }
    \caption{
    Examples of spectral energy distributions (SEDs) from COSMOS2020 galaxies alongside the best model solutions generated through Mephisto's chain of reasoning. We showcase three diverse galaxies to demonstrate Mephisto's capability to find solutions across a wide range of typical galaxy types: (a) A dusty star-forming galaxy with high attenuation ($\mathrm{A}_\mathrm{V}=2.84$) and active star formation, (b) a dwarf galaxy with low stellar mass and minimal dust, and (c) a massive galaxy potentially hosting an Active Galactic Nucleus component. The inset images show composite photometric images from the \href{https://www.legacysurvey.org/viewer}{Legacy Surveys Sky Viewer}.  Different colored lines represent contributions from different physical components (stellar emission in orange dashed lines, dust emission in purple dash-dotted lines, nebular emission in gray dotted lines, and AGN contribution in blue dash-dot ted lines where applicable), which combine to form the best-fit model (solid black line). Red data points with error bars show the observed fluxes across different wavelengths. Key derived physical parameters are displayed in the upper left of each panel, including stellar mass, dust attenuation, and star formation rate.
    }
    \label{fig:cosmos_demo}
\end{figure*}

Mephisto's effectiveness stems from two complementary memory systems, each addressing specific challenges in LLM-based scientific reasoning. For the current analysis, temporal memory maintains a compressed record of attempted model modifications and their outcomes, enabling Mephisto to avoid repeated mistakes and prioritize promising unexplored directions \cite[][]{FISCHER2023,ZHANGZEYU2024}. Given that each reasoning step requires approximately 15,000 tokens of context, this history is efficiently compressed using the LLM itself, preserving insights while managing context length.

Separately, an external knowledge base accumulates validated insights from previous analysis sessions, complementing the LLM's general capabilities with domain-specific information about SED modeling and CIGALE's functionalities. This knowledge undergoes a two-phase verification process: first, candidate insights are distilled from successful fitting sequences (e.g., ``when NIR bands show significant underestimation, exploring higher dust temperatures in the AGN module often improves the fit"; additional examples in Appendix~\ref{appendix:knowledge}); then, these candidates are validated through controlled experiments, running analyses both with and without the proposed knowledge. Only insights that demonstrably improve Mephisto's performance are retained and integrated into the knowledge base \cite[][]{VOYAGER2023,ZHAO2023,QIAN2024}, ensuring that accumulated expertise remains reliable and coherent.

\section{Results}\label{sec:result}

To demonstrate Mephisto's capabilities on properly reasoning the physical scenarios of galaxies through SED modeling, we design experiments on two scenarios: (1) reasoning on diverse, normal-type galaxies from the COSMOS2020 catalog \cite[][]{WEAVER2022} and (2) SED modeling on cutting-edge research frontiers -- ``Little Red Dots" galaxies revealed by James Webb Space Telescope \cite[][]{LABBE2023,PABLO2024}. We further show the temporal memory of Mephisto can facilitate efficient exploration, and Mephisto can learn from its self-play experience and distill helpful insights to improve its proficiency in SED modeling continually. We evaluate Mephisto's performance varying the large language model backbones to investigate whether open-source large language models can achieve reasonable performance to support large-scale deployment. During the main experiment, GPT-4o was adopted as the reasoning backbone. 

\subsection{SED Fitting of COSMOS2020 Galaxies}\label{subsec:overall}

To demonstrate Mephisto's general capability in SED modeling, we conducted experiments using galaxies from the COSMOS2020 catalog \cite[][]{WEAVER2022}, which represents one of the most detailed multi-wavenlength SED data for galaxies across a broad redshift range in a well-studied extragalactic field. Focusing on galaxies with reliable redshift measurements, we cross-matched COSMOS2020 with spectroscopic surveys including PRIMUS \cite[][]{PRIMUS2013}, 3D-HST \cite[][]{BRAMMER2012}, SDSS DR16 \cite[][]{SDSSDR162020}, C3R2 \cite[][]{C3R22019}, DEIMOS \cite[][]{DEIMOS2018}, MOSDEF \cite[][]{KRIEK2015}, and LEGA-C \cite[][]{VANDERWEL2021}, resulting in 38,547 sources from 723,897 observations.

To ensure an unbiased evaluation of Mephisto's performance across all galaxy types, rather than just the dominant classes in the catalog, we employed neural spline flow \cite[][]{DURKAN2019}, a neural network-based density estimation technique, in a four-dimensional space encompassing redshift, stellar mass, V-band dust attenuation, and star formation rate. We assigned weights inversely proportional to the estimated density and uniformly sampled 256 galaxies based on these weights. This approach resulted in a largely uniform sample spanning redshifts from 0.1 to 5.7, stellar masses between $10^6$ and $10^{12}\,\mathrm{M}_{\odot}$, dust attenuation values from 0 to 3, and star formation rates ranging from $10^{-4}$ to $10^{4}\,\mathrm{M}_{\odot}/\mathrm{yr}$, each with 30–40 photometric measurements at different wavelengths. We show the distributions of these key properties of the selected sample in Appendix~\ref{appendix:cosmos_data}.

Computationally intensive grid searches or Bayesian sampling—employed by established tools like CIGALE \cite[][]{CIGALE2019}, Prospector \cite[][]{JOHN2021}, and BEAGLE \cite[][]{CHEV2016}—have become standard practice in astronomical research to ensure well-sampled posterior distributions for galaxy properties. Thoroughly exploring complex, high-dimensional parameter spaces to avoid local minima and properly characterize uncertainties. However, these approaches either require exploring a pre-compiled large model grid consisting of hundreds of millions of models or employing a sophisticated sampling scheme and generating SED models on the fly, making them computationally expensive and limiting their application to the large sample size from future imaging surveys.

We aim to demonstrate that Mephisto can achieve comparable or superior results to these state-of-the-art approaches through iterative reasoning rather than brute force. While traditional methods rely on exhaustive searches across massive parameter spaces, Mephisto progressively refines its solutions through targeted exploration guided by physical understanding. Instead of calculating hundreds of millions of models simultaneously, Mephisto identifies directions and systematically improves its models through logical inference, allowing it to achieve high-quality fits while exploring only a fraction of the parameter space.

\begin{figure}
    \centering
    \includegraphics[width=0.8\linewidth]{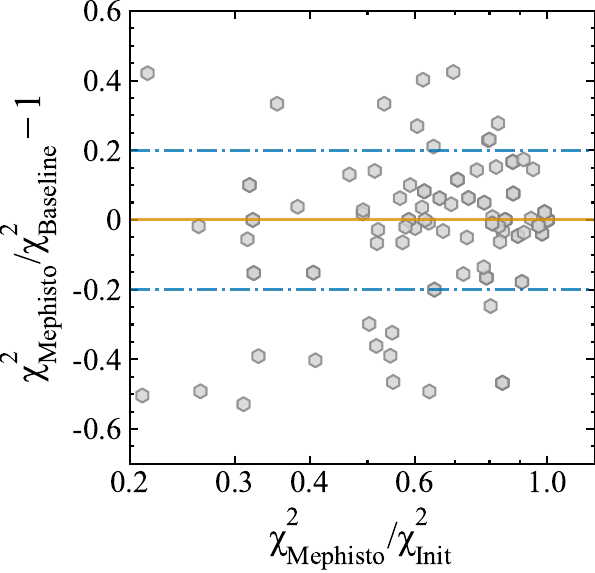}
    \caption{ 
Comparison of relative $\chi^2$ values for the 256 uniformly selected galaxies from the COSMOS2020 catalog, showing solutions found by Mephisto versus those obtained from exhaustive grid search models. The y-axis shows the fractional difference between Mephisto and exhaustive search $\chi^2$ values (($\chi^2_\text{Mephisto} - \chi^2_\text{Baseline}$)/$\chi^2_\text{Baseline}$), while the x-axis shows the normalized exhaustive search $\chi^2$ values relative to a initial SED model at the start of Mephisto's search ($\chi^2_\text{Baseline}/\chi^2_\text{Init}$). The exhaustive search (baseline model) represents our approach with 360 million grid points, while the basic fit refers to a simplified initial fit using standard templates without refinement. The blue dashed lines indicate the $\pm20$\% boundary, demonstrating that Mephisto consistently identifies solutions with $\chi^2$ values within 20\% of the exhaustive search, despite utilizing a grid size approximately 100 times smaller. Notably, many points fall below the zero line (orange), indicating cases where Mephisto produces better fits than the exhaustive search, while requiring fewer computational resources compared to traditional exhaustive searches.
}
    \label{fig:chi2compare}
\end{figure}

\begin{figure*}
    \centering
    \begin{tabular}{cc}
    \includegraphics[width=0.43\linewidth]{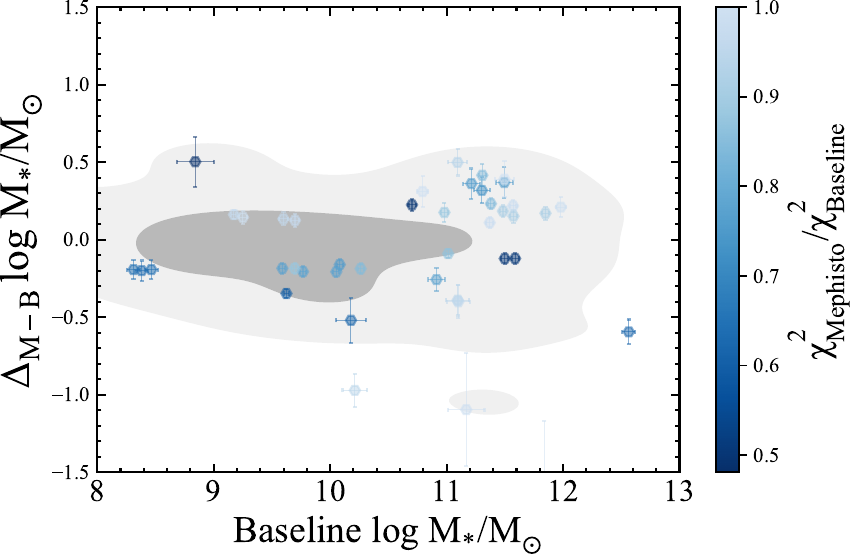} &
    \includegraphics[width=0.43\linewidth]{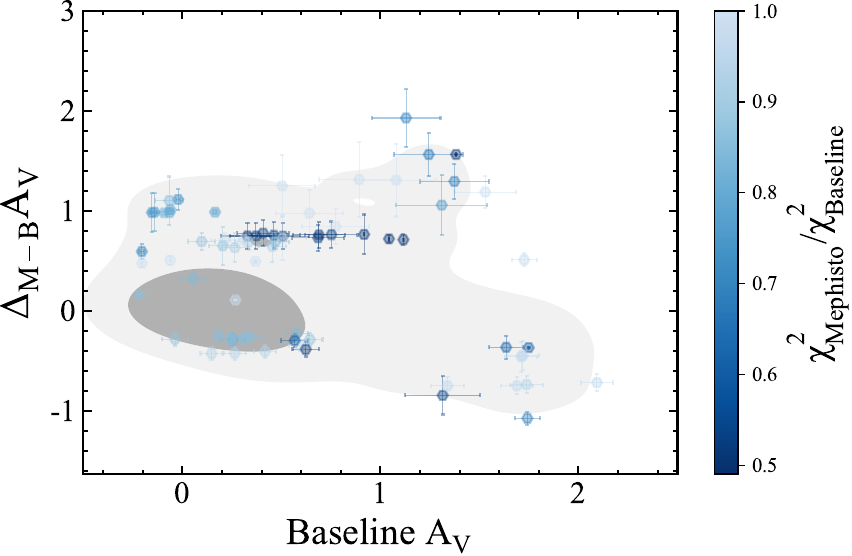} \\
    \end{tabular}
    \caption{Comparison of key physical parameter estimates between Mephisto and baseline grid search for 256 COSMOS2020 galaxies. The y-axis shows the difference between Mephisto and baseline estimates for the parameter in question, while the x-axis shows the baseline values. Color indicates the ratio of Mephisto's $\chi^2$ to the baseline model's $\chi^2$, with darker points representing superior fits by Mephisto. Points with error bars highlight cases where Mephisto finds better solutions ($\chi^2_{\text{Mephisto}}/\chi^2_{\text{Baseline}} < 1$). Contours illustrate the density distribution of solutions of the 256 galaxies. \textbf{Left panel:} Stellar mass estimates show strong consistency across the full mass range ($10^8$-$10^{12} {\rm M}_\odot$), with typical differences contained within $\pm$0.3 dex. Mephisto achieves equal or better constraints for most galaxies despite evaluating only $\sim$1\% of the parameter space, with greatest improvements (darker points with error bars) often seen at the extremes of the mass distribution. \textbf{Right panel:} Dust attenuation ($A_V$) estimates reveal Mephisto's ability to navigate parameter degeneracies. Despite $A_V$ being difficult to constrain due to its entanglement with stellar population, AGN contribution, and star formation history, Mephisto provides comparable or improved precision across the full range of attenuation values. The systematic improvement for high-attenuation systems ($A_V > 1$) demonstrates Mephisto's effectiveness in handling the complex physical scenarios that characterize dusty galaxies.}
    \label{fig:compare_params}
\end{figure*}

To establish a fair baseline for comparison, we implemented an exhaustive grid search using CIGALE with a carefully curated SED grid based on literature standards. This baseline grid follows the approach used in \cite{BROWN2014}, originally designed for 129 local galaxies with diverse characteristics. To better align the model with our sample's broad redshift range (0.1 to 5.7), we adopted a finer grid for the age of the main stellar population, extending it from 1 Gyr to 13 Gyr, and expanded the AGN emission fraction in the dust emission model to cover values from 0 to 0.99. The final baseline model comprises (detailed at Appendix~\ref{appendix:baseline}) a grid with 360 million points spanning 10 physical parameters—an approach typical of state-of-the-art SED modeling studies.

Throughout its reasoning process, Mephisto generates and evaluates numerous model configurations at each step. In total, Mephisto explores approximately 36 distinct reasoning steps for each galaxy, each involving the evaluation of multiple parameter combinations. For each direction identified during reasoning, Mephisto implements a ``zoom-in'' search strategy that dynamically determines appropriate parameter ranges to explore in greater detail. This targeted approach allows Mephisto to focus computational resources on the most relevant regions of parameter space, with the grid size and resolution determined by Mephisto's understanding of the specific galaxy's characteristics. We considered states with a relative $\chi^2$ within 10\% of the best solution among these reasoning paths. The states explored by Mephisto typically have 8 to 12 free parameters and a grid size ranging from $10^4$ to $10^6$ per step. For the entire reasoning process, the total grid size calculated ranges from $1 \times 10^6$ to $6 \times 10^6$, which is approximately 1\% of the baseline model's grid size.

Examples of galaxies within the catalog and their corresponding best CIGALE solutions by Mephisto are shown in Figure~\ref{fig:cosmos_demo}, which includes three different types: (a) a dusty star-forming galaxy, (b) a dwarf galaxy, and (c) a massive galaxy potentially hosting an Active Galactic Nucleus (AGN).

Despite requiring less computational resources, Mephisto's final states demonstrate consistency with the baseline model, as shown in Figure~\ref{fig:chi2compare}. The relative $\chi^2$ values obtained by Mephisto remain consistently within $\pm$20\% of those from the exhaustive grid search. In many cases, Mephisto actually identifies solutions that outperform the baseline model. Further analysis reveals that these superior states typically result from Mephisto's ability to adapt key components—particularly in star-formation history, dust, and AGN models—beyond the fixed assumptions inherent in the baseline exhaustive search. By dynamically tailoring solutions within the CIGALE framework rather than being constrained by predefined parameter grids, Mephisto achieves better fits while evaluating only about 1\% of the models required by the baseline approach.

To better understand the cases where Mephisto identifies superior solutions compared to traditional methods, we analyze parameter differences in Figure~\ref{fig:compare_params}. The figure compares Mephisto and baseline search results for two key physical parameters: stellar mass (left panel) and dust attenuation (right panel). The y-axis displays the difference in derived parameters (Mephisto minus baseline), while the x-axis shows the parameter value from the baseline. Points are color-coded by the relative improvement in $\chi^2$ values, with darker colors indicating where Mephisto found better fits. Points with error bars highlight cases where Mephisto finds better solutions ($\chi^2_{\text{Mephisto}}/\chi^2_{\text{Baseline}} < 1$).

Beyond computational efficiency, Mephisto's iterative approach offers advantages in constraining these key physical parameters. As shown in the left panel, stellar mass estimates—typically considered more robust in SED modeling—show strong consistency across the whole mass range ($10^8$-$10^{12} \mathrm{M}_\odot$), with typical differences contained within $\pm$0.3 dex. Mephisto achieves equal or better constraints for most galaxies despite evaluating only $\sim$1\% of the parameter space. In cases where Mephisto achieves better $\chi^2$ values (indicated by darker colors), it often derives slightly different stellar masses, with the most improvements (darker points with error bars) often seen at the extremes of the mass distribution. This enhanced precision stems from Mephisto's targeted exploration of AGN and stellar component interactions, which allows it to better account for their combined effects on the observed spectrum.

The right panel demonstrates that dust attenuation ($\mathrm{A_V}$)—a degenerate parameter entangled with stellar populations, star formation history, and AGN contributions—shows more variation between the methods but still maintains comparable or improved precision across the full range of attenuation values. The most considerable improvements in model fits (indicated by darker colors) often correspond to cases where Mephisto derives different dust attenuation values, particularly for high-attenuation systems ($\mathrm{A_V > 1}$). This demonstrates Mephisto's effectiveness in handling the complex physical scenarios that characterize dusty galaxies. Rather than blindly sampling a predefined parameter grid, Mephisto systematically isolates and tests specific physical assumptions, identifying which components most strongly influence the observational data. This approach allows it to navigate parameter degeneracies more effectively, finding optimal combinations of physical parameters that might be missed in grid-based searches.

\begin{figure*}[ht]
    \centering
    \gridline{
    \fig{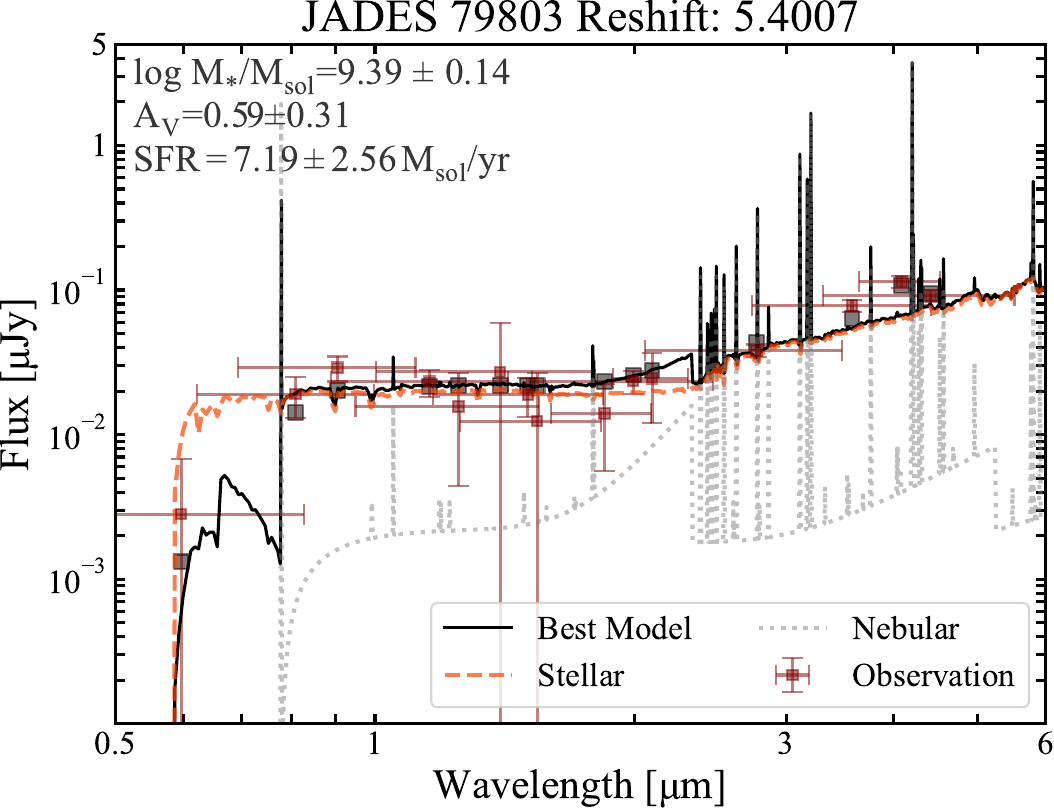}{0.4\textwidth}{JADES LRD 79803 Stellar Only Model\label{fig:lrd_demo_a}}
    \fig{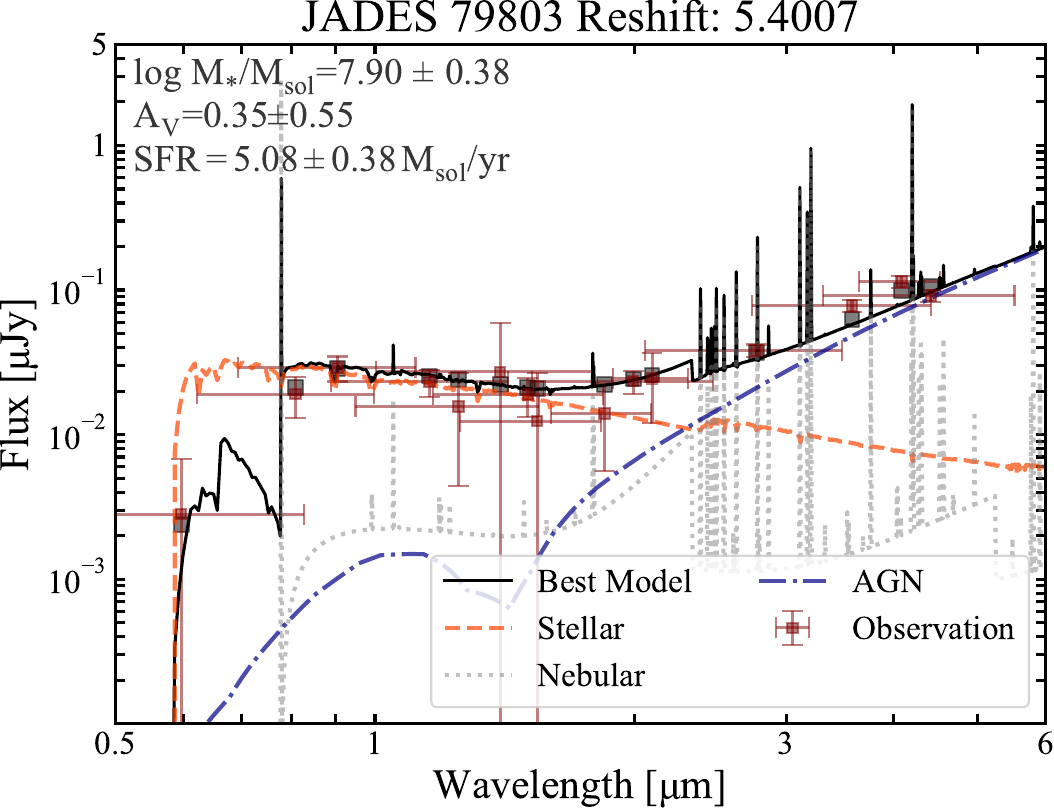}{0.4\textwidth}{JADES LRD 79803 Stellar + AGN model\label{fig:lrd_demo_b}}
    }
    \gridline{
    \fig{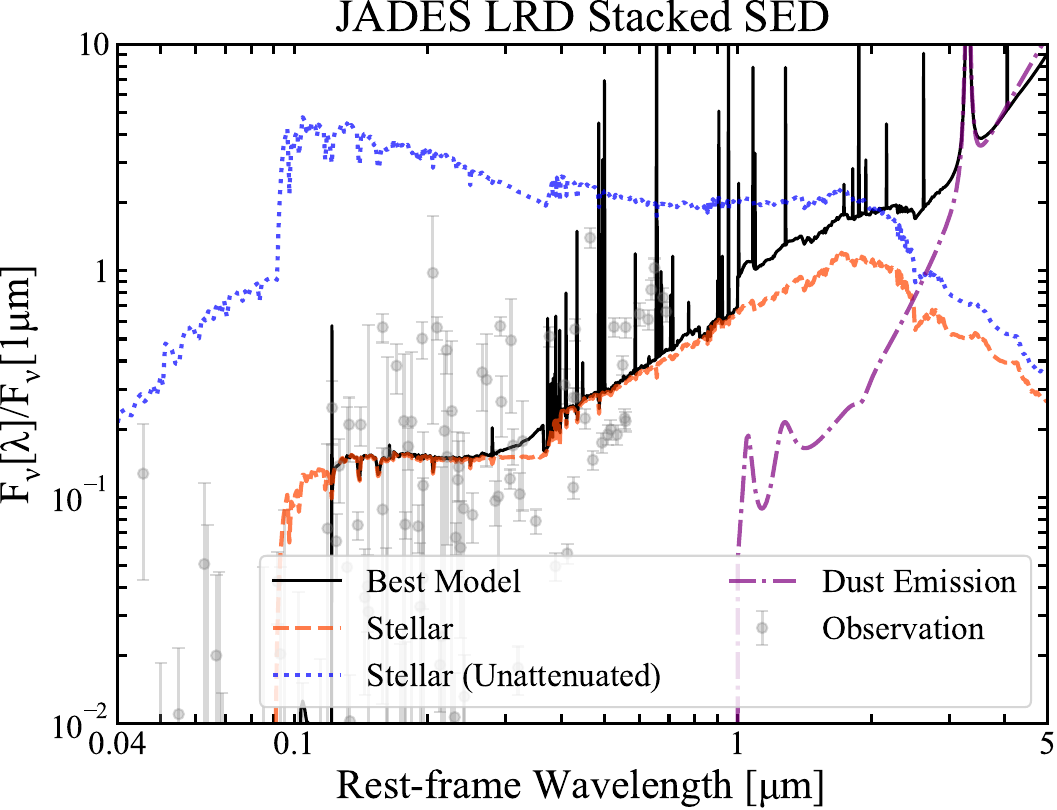}{0.4\textwidth}{Composite SED for 31 LRDs with Stellar Only Component\label{fig:lrd_composite_a}}
    \fig{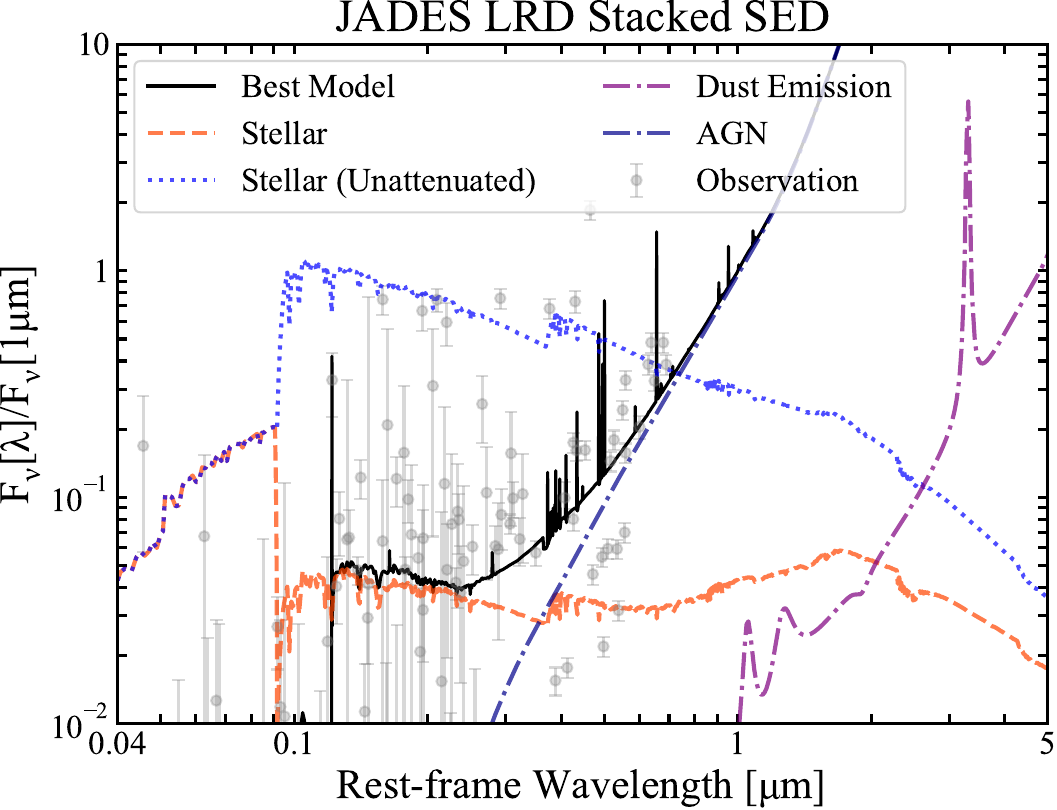}{0.4\textwidth}{Composite SED for 31 LRDs with Stellar + AGN Component\label{fig:lrd_composite_b}}
    }
    \caption{Mephisto's SED modeling of Little Red Dots. Top panels (a,b) show individual fits for a representative LRD (JADES 79803) using two different physical scenarios: (a) a stellar-only model representing a dust-obscured starburst galaxy, where the absence of an AGN contribution line is clear, and (b) a combined stellar+AGN model where the blue dash-dotted line shows the AGN contribution at longer wavelengths. Both scenarios lead to reasonable fits with similar overall quality ($\chi^2$ values) despite employing drastically different physical components and yielding different derived physical parameters (stellar mass, dust attenuation, and star formation rate as shown in the upper left corners). Bottom panels (c,d) display the composite SEDs for all 31 LRDs under both modeling approaches: (c) stellar-only and (d) stellar+AGN. For these composite plots, individual galaxy fluxes were normalized before stacking to identify common spectral features across the population. The characteristic V-shaped feature is evident in the rest-frame UV to optical wavelength range. Observations are shown as gray points with error bars, while different model components are represented by colored lines.}
    \label{fig:lrd_basic}
\end{figure*}

\begin{figure*}[htbp]
    \centering
    \gridline{
    \fig{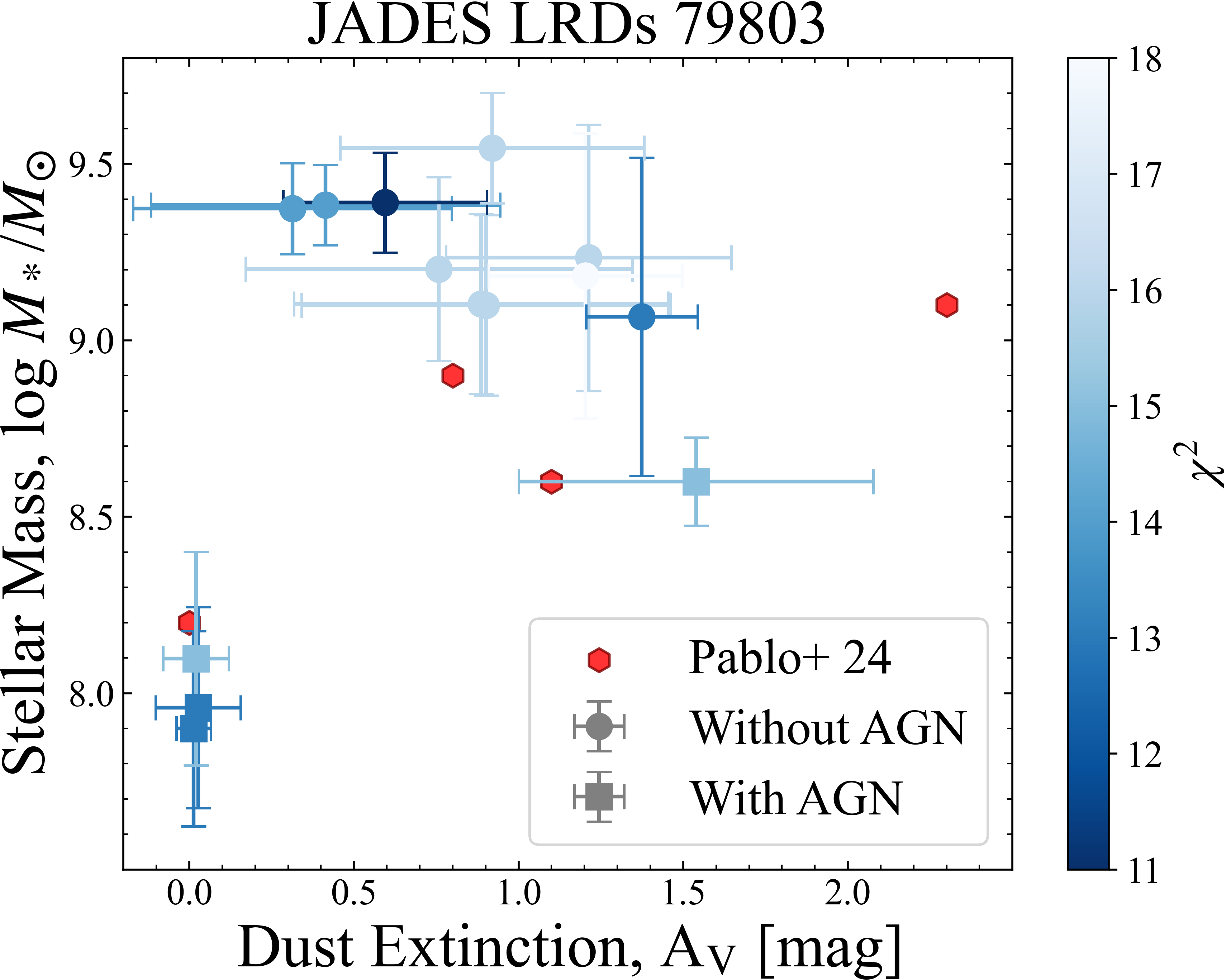}{0.43\textwidth}{Solution Space for JADES LRD 79803\label{fig:lrd_single}}
    \fig{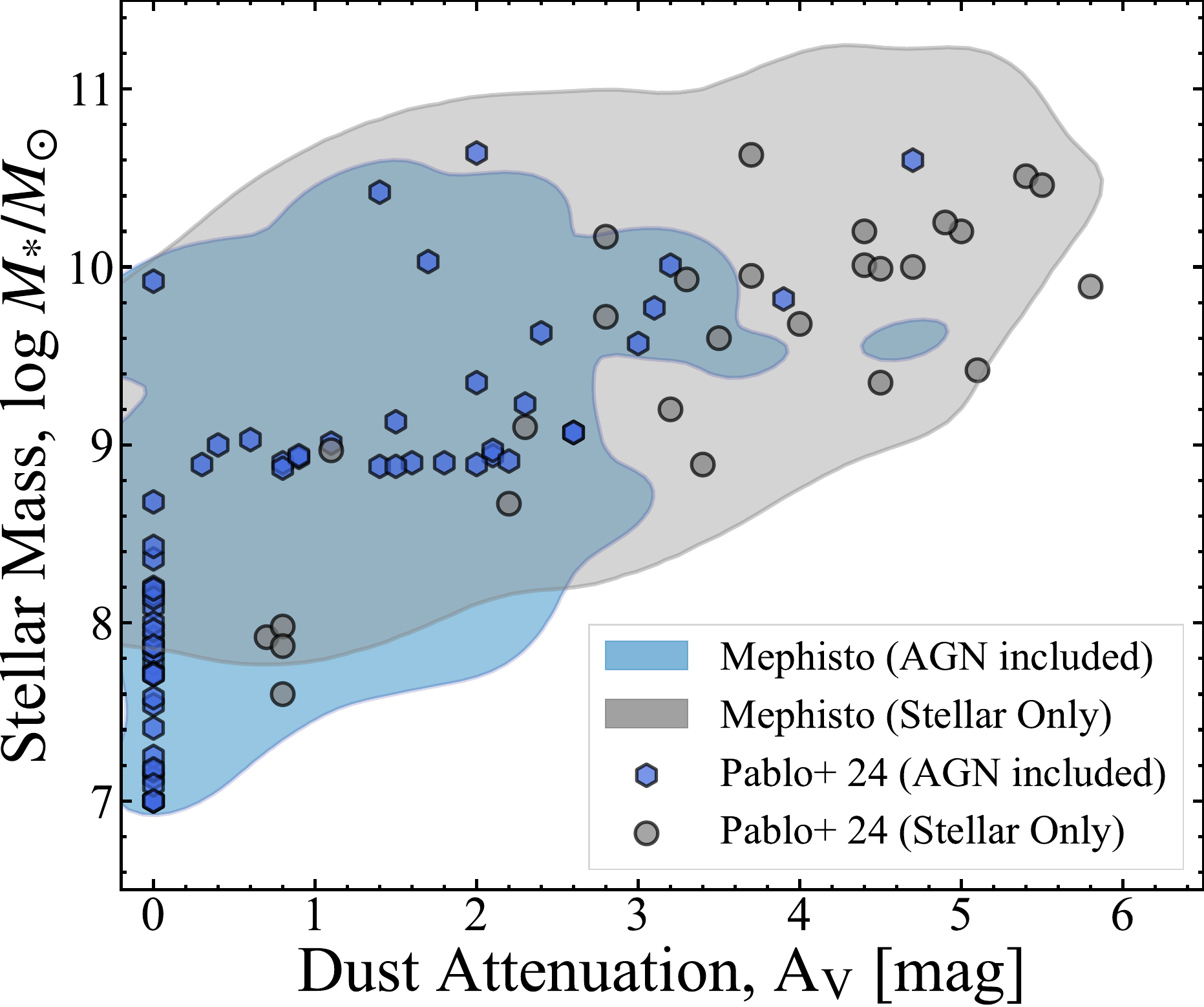}{0.39\linewidth}{Population Solution Space\label{fig:lrd_pop}}
    }
    \caption{Solution space analysis for Little Red Dots. (a) Parameter space exploration for a single LRD (JADES 79803), showing the relationship between stellar mass and dust extinction. Blue points represent Mephisto's solutions, with darker shades indicating better fits (lower $\chi^2$ values). The solutions clearly separate into two groups: lower-extinction models without AGN components (bottom cluster) and higher-extinction models with AGN components (top cluster). Red points show literature solutions from \cite{PABLO2024}, which align well with Mephisto's findings. (b) Population-level comparison of physical parameters derived by Mephisto (blue \& grey density plot) and \cite{PABLO2024} (blue hex and grey dots) for all 31 LRDs.}
    \label{fig:lrd_solution_space}
\end{figure*}

\subsection{Little Red Dots}\label{subsec:LRD}

While the preceding results demonstrate Mephisto's effectiveness with conventional galaxies—even surpassing exhaustive baseline approaches in certain cases—the true value of agentic scientific reasoning emerges when confronting phenomena that fall outside established domains. The most scientific discoveries often involve objects or processes that defy existing models and expectations, requiring exploratory analysis rather than mere optimization within known parameter spaces.

To evaluate Mephisto's capability in addressing astrophysical problems, we tested it on the enigmatic ``Little Red Dots'' (LRDs) using the James Webb Space Telescope (JWST) dataset from \cite{PABLO2024}, which includes 31 LRDs. These compact, optically red high-redshift galaxies exhibit a point-like morphology and a distinctive V-shaped spectral feature spanning ultraviolet to optical wavelengths—a characteristic that challenges conventional SED models of low-redshift galaxies. The nature of LRDs remains debated, though two leading hypotheses dominate current discourse. The first posits that LRDs host obscured active galactic nuclei (AGN), where dust-reddened light originates from accreting supermassive black holes \cite[][]{MATT2024, PABLO2024, GREENE2024}. The second suggests they are dust-obscured starburst galaxies, where intense star formation is shrouded by dense interstellar material that reddens stellar radiation through absorption \cite[][]{XIAO2023, GENTILE2024, BAGGEN2024}. By applying Mephisto to these 31 LRDs, we assess its ability to disentangle these mechanisms and advance SED modeling in exploratory regimes.

For our experiments, we utilize photometric data from the JWST Advanced Deep Extragalactic Survey (JADES) Data Release 2 \cite[][]{JADES2023}, spanning 15–21 optical to mid-infrared bands. This dataset combines observations from JWST's Near-Infrared Camera (NIRCam) and the Hubble Space Telescope's (HST) Wide Field Camera 3 (WFC3). Redshift measurements for the sample were adopted directly from \cite{PABLO2024}, which include spectroscopically confirmed redshifts where available and photometric redshift estimates for the remaining sources. Mephisto employs a breadth-first search strategy, with the maximum number of explored states set to 36. The full catalog used for LRD analysis is available in the Supplementary Materials.

The top panels of Figure~\ref{fig:lrd_basic} demonstrate Mephisto's analysis of a representative LRD (JADES 79803). Through iterative reflection on discrepancies between SED models and observational data, Mephisto identifies two viable physical scenarios for this unusual object. For individual LRDs like JADES 79803, Mephisto independently evaluates both a dusty starburst interpretation (stellar-only model, Figure~\ref{fig:lrd_basic}a) and an obscured AGN hypothesis (stellar+AGN model, Figure~\ref{fig:lrd_basic}b), determining that either could provide reasonable fits to the observational data. The consistency of this dual interpretation extends across the entire sample, as shown in the composite SEDs (bottom panels). These stacked spectra combine normalized fluxes from all 31 LRDs to reveal common spectral features and systematic differences between the two modeling approaches (Figure~\ref{fig:lrd_basic}c and Figure~\ref{fig:lrd_basic}d). The composite analysis highlights that both solutions—starburst and AGN—remain consistently compatible with the observed SEDs across the population, explaining the ongoing vigorous discourse in the astronomical community regarding the true nature of these enigmatic systems.

Figure~\ref{fig:lrd_solution_space} presents a detailed analysis of the solution space for LRDs, comparing physical properties derived by Mephisto with those reported in the literature. For individual objects like JADES 79803 (Figure~\ref{fig:lrd_solution_space}a), Mephisto identifies two distinct solution clusters: low-extinction models without AGN and higher-extinction models with AGN components. These clusters align well with previously published results (red points), demonstrating Mephisto's ability to automatically identify physically viable interpretations at both individual and population levels. The relationship between stellar mass ($\mathrm{M}_{*}$) and V-band dust attenuation ($\mathrm{A}_\mathrm{V}$) derived by Mephisto across the entire sample of 31 LRDs (Figure~\ref{fig:lrd_solution_space}b) similarly shows good agreement with previous human expert analyses, confirming the robustness of Mephisto's approach. The consistency between Mephisto's result and the findings of multiple recent works on this topic \cite[][]{XIAO2023,LABBE2023,MATT2024,GENTILE2024,BAGGEN2024} validates Mephisto's ability to emulate the scientific reasoning process of domain experts, even when investigating phenomena at the boundaries of current understanding.

Although Mephisto can produce interpretations similar to those in early-stage LRD studies \cite[e.g.,][]{XIAO2023,LABBE2023,GENTILE2024}, it struggles to address more advanced hypotheses, such as those involving supermassive black holes embedded in dense gas environments \cite[][]{NAIDU2025,RUSAKOV2025,TAYLOR2025}. This limitation arises from two main factors.

First, Mephisto's current flexibility is limited: it lacks robust visual understanding and reasoning capabilities, which restricts advanced analysis based on visual features such as galaxy morphology and spectral characteristics. Furthermore, its action space is confined to CIGALE's predefined model framework and does not incorporate SED modeling components like CLOUDY \cite[][]{FERLAND2013} for nebular emission or FSPS \cite[][]{FSPS2009,FSPS2010} for stellar population synthesis. As a result, even if Mephisto can conceptually develop physical models beyond CIGALE's library, it cannot directly refine or test such hypotheses.

A more notable limitation lies in the current generation of LLMs, which exhibit restricted memory and reasoning capacities in open-world settings. While they show improvements at closed-ended tasks—such as solving advanced mathematical problems—their ability to handle open-ended scientific questions remains underdeveloped. For instance, although certain forms of abstract reasoning (such as drawing analogies from stellar atmospheres to black hole atmospheres) may seem intuitive to humans, AI systems still struggle to develop such cognitive processes autonomously.

Therefore, in its current form, Mephisto functions more as an AI copilot rather than an autonomous AI scientist. Nevertheless, with the ongoing development of foundation models, the realization of a fully capable AI scientist remains possible.


\subsection{Mephisto as experiential learner}\label{subsec:learn}

Beyond its ability to reason about individual astronomical observations, Mephisto incorporates mechanisms that enable it to continuously improve through experience. Unlike conventional LLM applications, Mephisto's success stems from its ability to systematically reflect on past reasoning attempts, extract generalizable insights, and accumulate domain knowledge—capabilities that elevate its performance beyond what a basic LLM implementation could achieve. This section examines how temporal memory and knowledge distillation contribute to Mephisto's capacity for self-improvement through experimental ablation studies conducted on the COSMOS2020 catalog.

\begin{figure}
    \centering
    \includegraphics[width=0.9\linewidth]{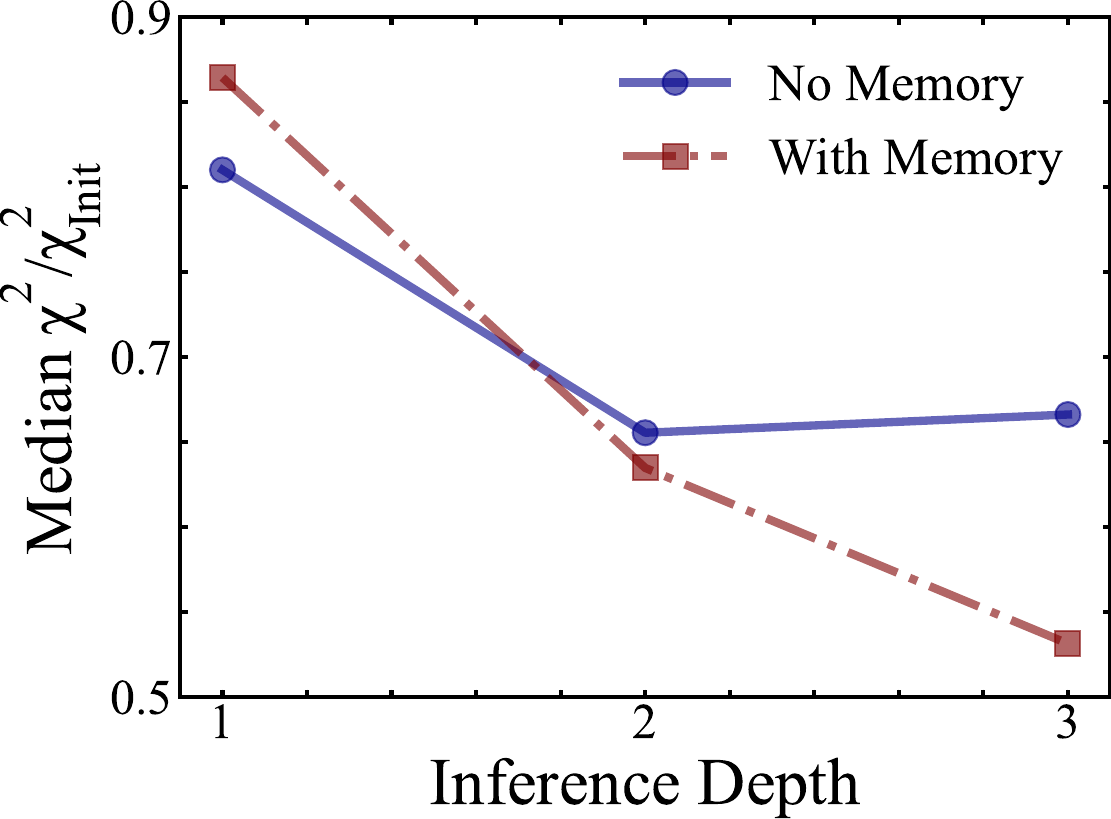}
    \caption{Comparative study demonstrating the impact of temporal memory on Mephisto's performance across inference depths. With memory enabled (red dashed line with squares), Mephisto shows consistent improvement at each depth, achieving a median relative $\chi^2$ (compared to the initial SED model at the start of Mephisto's search) of 0.52 by depth 3, compared to 0.67 without memory (blue solid line with circles)—a 25\% greater improvement. At the first depth, performance is comparable since both configurations start from identical states, but the memory-enabled system's ability to learn from previous attempts becomes increasingly apparent at depths 2 and 3. This confirms that temporal memory allows Mephisto to progressively refine its solutions by avoiding previously unsuccessful approaches and prioritizing promising directions.}
    \label{fig:ablation-study-memory}
\end{figure}

Temporal memory enables Mephisto to maintain a compressed record of previous reasoning attempts within the current analysis session, allowing it to avoid repeating unsuccessful approaches and to build upon promising directions. To assess its effectiveness, we compared Mephisto's performance with and without memory by manually disabling memory updates across all 256 COSMOS2020 galaxies. As illustrated in Figure~\ref{fig:ablation-study-memory}, both configurations show similar performance at the first inference depth, as expected since no previous attempts exist to reference at this initial stage. However, at the second and third inference depths, performance differences emerge. With memory enabled, Mephisto avoids making redundant failed attempts and instead focuses on refining approaches or exploring previously unexplored regions of the solution space.

This strategic refinement leads to a faster and more substantial decrease in relative $\chi^2$ (relative to the initial SED model at the start of Mephisto's search) during deeper inference stages, ultimately achieving a 25\% greater improvement by the third inference depth (median relative $\chi^2$ of 0.52 with memory versus 0.67 without memory). Figure~\ref{fig:ablation-study-memory} displays both the median relative $\chi^2$ values across all trials and galaxies, demonstrating that the performance advantage of memory-enabled reasoning is consistent, not merely the result of fortunate exploration in a few cases. This confirms that temporal memory enables Mephisto to leverage its past reasoning attempts to more effectively navigate the complex parameter space of SED models—a capability particularly valuable for addressing challenging galaxy observations that require multiple refinement iterations.

The knowledge base serves as Mephisto's long-term memory, accumulating validated insights across multiple analysis sessions and enabling transfer learning between different galaxy observations. To evaluate its effectiveness, we designed a controlled experiment using the COSMOS2020 catalog. We designated 48 galaxies as a test set and used the remaining 208 galaxies as a training set for knowledge accumulation. When analyzing the test galaxies, we deliberately prevented updates to the knowledge base, allowing us to assess whether knowledge gained from previously analyzed galaxies could improve performance on new, unseen sources.

Our knowledge distillation agent automatically extracted 72 detailed knowledge entries from interaction histories of the 208 galaxies in the training set. These entries focus on specific details of \texttt{CIGALE}'s physical models—information more precise than the general spectral energy distribution (SED) knowledge inherent to large language models. For example: ``If mid-infrared (3–25$\,\mu$m) flux is underestimated, increasing parameters for dust extinction (\texttt{EBV}) and active galactic nucleus (AGN) contribution (\texttt{fracAGN}) can improve model fits.'' This detailed knowledge—which human astronomers typically acquire through years of trial and error with different modeling approaches—is gradually acquired by Mephisto through systematic knowledge distillation from its own reasoning attempts. Detailed extracted knowledge and prompts for knowledge distillation are provided in the Supplementary Information.

\begin{figure}
    \centering
    \includegraphics[width=0.9\linewidth]{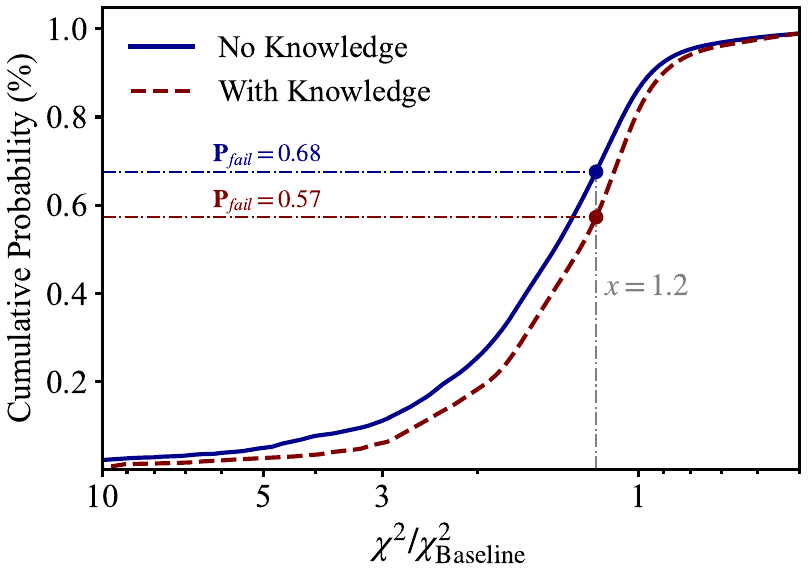}
    \caption{Ablation study of Mephisto's solution distributions with and without the distilled knowledge base on 48 COSMOS2020 galaxies. The knowledge is distilled automatically from the interaction history between Mephisto and the other 208 COSMOS2020 galaxies. The figure shows the cumulative probability distribution of the ratio between Mephisto's $\chi^2$ and the baseline $\chi^2$ values. The ``With Knowledge'' curve (red dashed line) consistently lies below the ``No Knowledge'' curve (blue solid line), demonstrating systematic improvement in solution quality. Using the failure criterion ($\chi^2/\chi^2_{\text{Baseline}} > 1.2$) as an illustrative example, the ``No Knowledge'' case has $\mathbf{P}_{\text{fail}} = 0.68$, while ``With Knowledge'' has $\mathbf{P}_{\text{fail}} = 0.57$, showing an 11\% improvement in generating valid solutions. This highlights how accumulated domain knowledge enhances Mephisto's exploration efficiency for previously unseen galaxies.}
    \label{fig:abaltion-study-knowledge}
\end{figure}

As illustrated in Figure~\ref{fig:abaltion-study-knowledge}, the utility of the knowledge base is evident when comparing solution distributions from Mephisto with versus without the knowledge base. The figure shows the cumulative probability distribution of the ratio between Mephisto's $\chi^2$ and the baseline $\chi^2$ values, with lower ratios indicating better performance. Across the entire distribution, the ``With Knowledge'' curve consistently lies below the ``No Knowledge'' curve, demonstrating a systematic improvement in solution quality across all performance thresholds. 

This shift in the cumulative distribution means that for any chosen quality threshold, Mephisto with knowledge consistently produces more high-quality solutions than without knowledge. For example, using a failure criterion where a state is deemed failed if $\chi^2/\chi^2_{\text{Baseline}} > 1.2$, we find that Mephisto exhibits an 11\% higher probability of producing physically meaningful interpretations when equipped with the distilled knowledge (failure probability of 0.57 with knowledge versus 0.68 without). The improvement is not limited to this specific threshold—the knowledge-equipped system shows advantages across a wide range of $\chi^2$ ratios. This demonstrates that the accumulated knowledge helps Mephisto navigate the solution space more effectively, allowing it to find not just marginally better solutions but substantially better ones in many cases.

In summary, our ablation studies demonstrate that the integration of temporal memory and knowledge distillation enhances Mephisto's capabilities in SED modeling by enabling it to learn from experience in complementary ways. Temporal memory provides short-term guidance within individual reasoning sessions, while the knowledge base captures longer-term insights that transfer across different astronomical objects. These mechanisms demonstrate how large language models can transcend their initial training to develop specialized expertise through structured self-improvement.

\subsection{Evaluation on different large language models}

While we have demonstrated Mephisto's ability to distill useful knowledge from exploration experience, the experiments discussed thus far have utilized GPT-4o as the LLM backbone. During the time this project was conducted (summer 2024), GPT-4o represented one of the state-of-the-art models, with proven instruction-following capabilities—a key criterion for this study. It also possessed substantial knowledge about astronomy through its training \cite[][]{TING2024}.

For practical applications in astronomy, however, cost-performance efficiency becomes important. With billions of sources discovered by ongoing and upcoming imaging surveys such as Roman \cite[][]{ROMAN2023}, Euclid \cite[][]{EUCLID2022}, and CSST \cite[][]{CSST2023}, scaling up with GPT-4o would be prohibitively expensive. At approximately two dollars per galaxy for detailed analysis requiring 200K tokens, performing SED fitting on all sources would cost billions of dollars—comparable to the construction budget of these space missions themselves. A more practical approach would be to extract knowledge using high-performance models like GPT-4o initially, then deploy this knowledge with more cost-effective models for large-scale analysis.

To explore this strategy, we conducted a comparative study using the same 256 COSMOS2020 galaxies described previously, while varying the LLM backbone\footnote{This exploration is non-exhaustive and reflects models accessible to the research team during early 2025.}. The evaluation encompassed seven different LLMs across three distinct cost tiers:
(1) \textbf{High-cost tier} (baseline): GPT-4o (relative cost: 1.0); (2) \textbf{Mid-cost tier} ($\sim$1/10 relative cost): DeepSeek-V3\footnote{\href{https://huggingface.co/deepseek-ai/DeepSeek-V3}{https://huggingface.co/deepseek-ai/DeepSeek-V3}}, DeepSeek-R1\footnote{\href{https://huggingface.co/deepseek-ai/DeepSeek-R1}{https://huggingface.co/deepseek-ai/DeepSeek-R1}}, QwQ-32B\footnote{\href{https://huggingface.co/Qwen/QwQ-32B}{https://huggingface.co/Qwen/QwQ-32B}}, DeepSeek-R1-Distill-Llama-70B\footnote{\href{https://huggingface.co/deepseek-ai/DeepSeek-R1-Distill-Llama-70B}{https://huggingface.co/deepseek-ai/DeepSeek-R1-Distill-Llama-70B}}; (3) \textbf{Low-cost tier} ($\sim$1/100 relative cost): DeepSeek-R1-Distill-Llama-8B\footnote{\href{https://huggingface.co/deepseek-ai/DeepSeek-R1-Distill-Llama-8B}{https://huggingface.co/deepseek-ai/DeepSeek-R1-Distill-Llama-8B}}, GLM-4-Air-250414\footnote{\href{https://open.bigmodel.cn/dev/howuse/model}{https://open.bigmodel.cn/dev/howuse/model}}

To ensure a fair comparison, each agent was provided with knowledge previously distilled from GPT-4o interactions, minimizing the impact of differences in internal domain knowledge among the LLMs. This approach reflects a practical deployment scenario: knowledge distillation would be conducted once with the best-performing model, and the resulting insights—expressed in natural language—can be shared with more affordable models for large-scale application. The SED fitting task requires the LLM to accurately interpret CIGALE's documentation, adhere to prompt instructions, input data, and effectively utilize provided knowledge and memory—all of which demand advanced instruction-following and long-context reasoning capabilities.

\begin{figure}[htbp]
    \centering
    \includegraphics[width=0.95\linewidth]{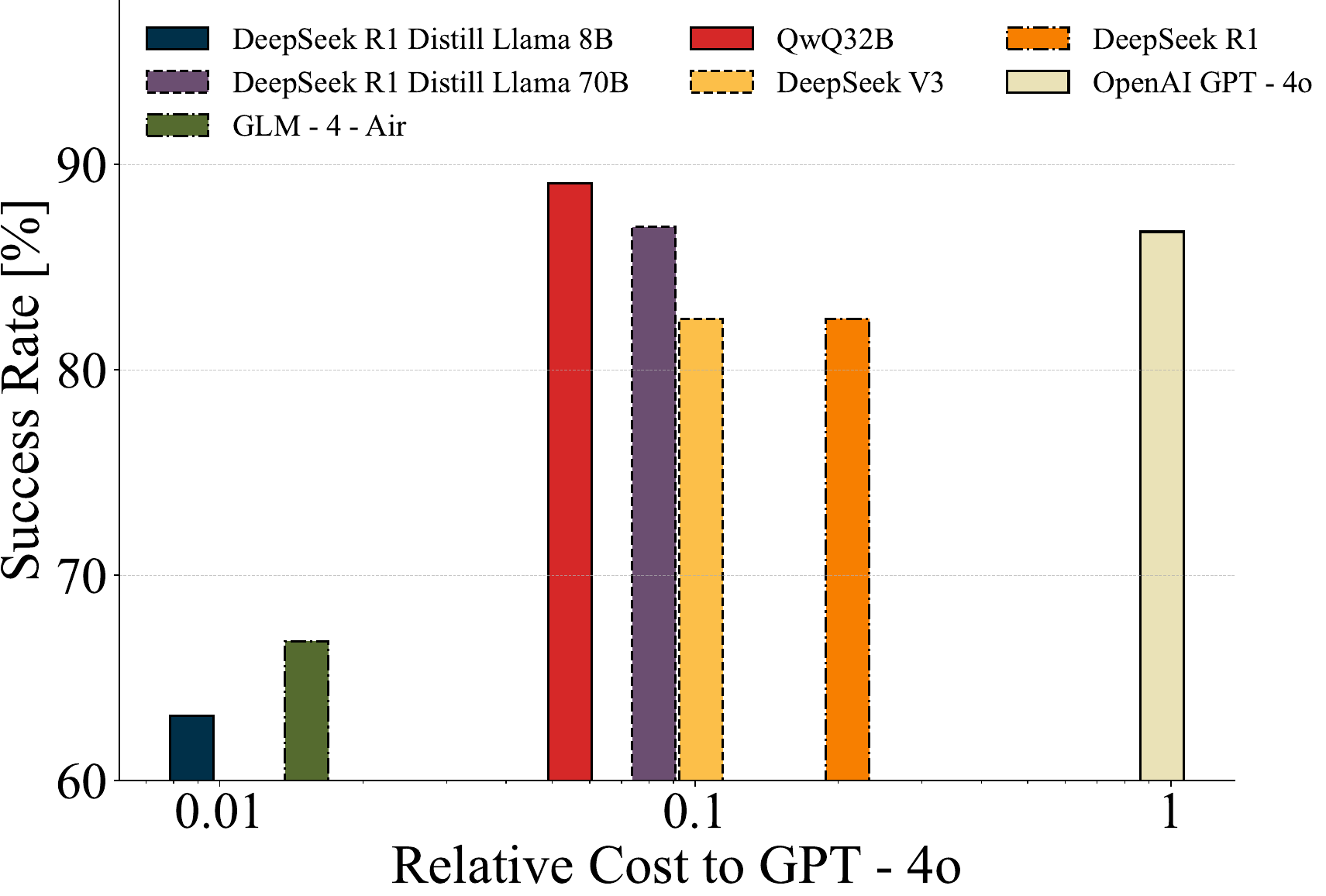}
    \caption{Cost-performance comparison of LLMs for Mephisto's SED fitting task across three relative cost tiers. The y-axis shows success rates (percentage of galaxies where Mephisto found at least one solution with $\chi^2 < 1.2\times\chi^2_\mathrm{baseline}$). Most models achieve 80-90\% success rates, with low-cost models ($\sim$1/100 relative cost: DeepSeek-R1-Distill-Llama-8B and GLM-4-Air) showing more notable performance drops. All models utilize knowledge pre-distilled using GPT-4o, with high-cost GPT-4o (relative cost = 1.0) serving as the baseline. This analysis demonstrates the viability of using cost-effective LLMs with pre-distilled knowledge for large-scale astronomical surveys,.}
\label{fig:diffllm}
\end{figure}

Figure~\ref{fig:diffllm} shows Mephisto's performance across different LLM backbones. We estimated costs by averaging the input and output token prices\footnote{LLM pricing data are derived as the median value from pricing information of various API service providers available at \href{https://artificialanalysis.ai/leaderboards/providers}{https://artificialanalysis.ai/leaderboards/providers} as of May 2025.} during the experimental period, presenting them relative to GPT-4o (set at 1.0).

For evaluation purposes, we define that Mephisto provides a reasonable interpretation if it generates any SED model yielding a solution with $\chi^2 < 1.2\times\chi^2_\mathrm{Baseline}$, following our findings in Figure~\ref{fig:chi2compare}. It's important to note that while the exhaustive search represents an expensive grid search across hundreds of millions of possible parameter combinations, this comparison metric primarily illustrates how well Mephisto can perform with just 1\% of the computational cost. As shown in our earlier analysis, Mephisto can sometimes even perform better than the exhaustive search by more effectively exploring the parameter space.

The performance analysis revealed that most models achieve impressive success rates of 80-90\% by this metric. Low-cost models (DeepSeek-R1-Distill-Llama-8B, GLM-4-Air-250414) showed $>$25\% success rate decreases compared to GPT-4o, though they still provide reasonable results for many galaxies. There are performance variations among the mid-cost models, with QwQ-32B, DeepSeek-R1, and DeepSeek-V3 all performing well relative to GPT-4o. These variations should not be over-interpreted, as the metric itself is not based on ground truth but serves as a performance comparison to assess the viability of different models when paired with knowledge pre-distilled from GPT-4o. Overall, mid- and high-cost tiers effectively supported the SED modeling task with ``success'' rates consistently above 80\%.

These findings have important implications for large-scale astronomical surveys. While using GPT-4o for analyzing all sources in missions like Roman or Euclid would be prohibitively expensive, using mid-tier models like QwQ-32B would reduce computational costs to approximately 5-10\% of the mission's construction budget—a much more reasonable allocation for data analysis. This makes SED fitting for billions of sources financially viable, substantially increasing the scientific return on investment for these missions.

This analysis demonstrates that Mephisto can operate effectively with cost-effective LLMs beyond GPT-4o, provided pre-distilled domain knowledge is leveraged. While low-cost models currently lag in performance, mid-cost alternatives like QwQ-32B offer  tradeoffs, aligning with the trajectory of LLMs toward improving performance and reducing costs. As LLMs evolve, agentic frameworks like Mephisto become increasingly viable for large-scale astrophysical analysis, enabling efficient exploration of billions of astronomical sources to maximize the scientific return of next-generation surveys. 


\section{Discussion}\label{sec:discussion}

Modern astronomical surveys generate petabyte-scale datasets daily, yet human astronomers can analyze only a tiny fraction of these observations. This bottleneck is not merely computational—it reflects the challenge of scientific interpretation. While AI has automated many technical tasks in astronomy, from object classification to simulation acceleration, the core intellectual work of hypothesis generation and model selection remains firmly in human hands. This limitation becomes increasingly problematic as survey volumes grow: how many discoveries lie hidden in unexplored data, waiting for an astronomer who will never have time to look?

Mephisto represents a different approach to this challenge. Rather than treating astronomical analysis as a pattern recognition problem, it emulates the actual scientific process—proposing hypotheses, testing models, and refining interpretations based on evidence. Unlike black-box machine learning methods that map inputs to outputs without explanation, Mephisto provides transparent, step-by-step reasoning that astronomers can examine, validate, and build upon. This transparency is necessary for scientific credibility and enables the system to tackle phenomena outside its training data, as we demonstrated with the Little Red Dots that were discovered after most LLMs' knowledge cutoff.

The choice of spectral energy distribution (SED) fitting as our initial focus was deliberate. SED analysis encapsulates the essence of astrophysical reasoning: astronomers must identify which physical mechanisms—star formation, dust obscuration, active galactic nuclei—best explain observed multi-band photometry. This requires navigating a complex model space where different combinations of components can produce similar observations, making interpretation as important as optimization. Like medical diagnosis, success depends not just on finding a mathematical fit but on understanding which physical scenario makes sense given all available evidence. This makes SED fitting an ideal testing ground for agent-based astronomical research—challenging enough to be meaningful, yet structured enough to be tractable.

By interacting with the CIGALE codebase and its library of physical models, Mephisto learns to navigate the complex solution space efficiently. Our ablation studies confirm that both temporal memory (allowing the agent to avoid repeating unsuccessful approaches) and knowledge accumulation (transferring insights gained from previous analyses) are important components of this learning process. The experimental results clearly demonstrate that Mephisto develops a kind of analytical ``intuition'' through experience, similar to how human researchers refine their hypotheses and strategies.

We demonstrated two key capabilities of Mephisto. First, for normal galaxies from the COSMOS2020 catalog, we showed that Mephisto can achieve results comparable to those obtained from exhaustive grid searches, but with two orders of magnitude less computational effort. The agent learns to take ``shorter path", identifying the most relevant physical components and parameter ranges without brute-forcing the entire solution space. Second, and perhaps more interestingly, Mephisto successfully analyzes ``unknown unknowns''—the enigmatic Little Red Dots recently discovered by the James Webb Space Telescope. Despite these objects being discovered after most LLMs' knowledge cutoff dates, Mephisto independently arrived at the two leading hypotheses currently debated in the astronomical community: obscured active galactic nuclei and dust-obscured starburst galaxies. This ability to reason about and provide plausible interpretations for novel phenomena highlights an advantage over traditional machine learning approaches, which typically struggle with data that lies outside their training distribution.

An advantage of Mephisto, and one that distinguishes it from many ``black-box'' AI methods, is the complete traceability of its reasoning processes. Unlike models that directly map observations to physical properties, Mephisto provides a step-by-step account of its analysis, justifying its choices and explicitly revealing the underlying physical models it considers. This transparency is not just a desirable feature; it's needed for scientific rigor and allows seamless integration of Mephisto into the prevailing research paradigm. Human experts can scrutinize Mephisto's reasoning chain, validate its conclusions, and even use its output to guide further theoretical development or observational follow-up.

The implications for future astronomical surveys can be important. Ongoing and upcoming large-scale imaging surveys will gradually provide  deliver detailed coverage of billions of galaxies across multiple wavelengths, spanning from X-ray to radio. In this context, Mephisto presents a new approach to analyzing such vast datasets in a transparent, human-interpretable manner. By systematically exploring diverse physical interpretations of observational data, Mephisto could help identify sources that cannot be explained by current physical models, potentially leading to the discovery of new astrophysical phenomena that might otherwise evade detection through standard triage procedures.

However, this study also underscores several  limitations that must be addressed in future endeavors. The most obvious one is cost. The current implementation of Mephisto, using GPT-4o as its LLM backbone, costs roughly 2 US dollars per source for a detailed analysis. While this is acceptable for targeted studies of specific galaxy populations, it's prohibitively expensive for analyzing billions of sources. In fact, analyzing the entire dataset from missions like Euclid or Roman at this cost would be comparable to the build cost of the missions themselves (on the order of a billion dollars). 

Yet, there is ample reason for optimism. As demonstrated in \cite{TING2024}, the cost-effectiveness of LLM reasoning has improved in a very short time. Our experiments with alternative LLMs (Figure \ref{fig:diffllm}) suggest that comparable performance might be achievable at a fraction of the cost, particularly with models like QwQ-32B onpar success rates at just 5\% of the cost of GPT-4o. Even during the course of this project, we've seen models emerge that are orders of magnitude cheaper, potentially reducing the analysis cost to a small percentage of the mission cost. A viable strategy involves using high-performance models like GPT-4o for initial knowledge distillation, and then deploying this accumulated knowledge with more affordable models for large-scale analysis.

Perhaps a more important limitation, and one that applies to all current LLM-based agents, is the inherent capability of the foundation models themselves. Even with the reasoning abilities of models like GPT-4o, current vision-language models are still highly limited in their ability to understand even relatively simple scientific plots \cite[e.g.,][]{CHARXIV2024,YUWEIYANG2025}. This is a major reason why we restricted our current study to SED modeling, where the limited number of wavelength data points could be effectively ``textualized.'' This allowed us to leverage the reasoning capabilities of LLMs to efficiently navigate the vast hypothesis space, circumventing the visual processing limitations that remain a bottleneck for current vision-language models. While fine-tuning existing large vision-language models on specific tasks using scientific datasets \cite[][]{YUE2023,LILEI2024} offers a partial mitigation, further advancements in this area are needed.

Beyond the LLMs themselves, the development of AI agents for astronomical tools necessitates a unified data platform that enables seamless access to, and manipulation of, diverse datasets. Modern astronomical research often requires integrating data from multiple surveys, and these data are frequently stored in heterogeneous, and sometimes archaic, formats. In this study, we benefited from the meticulously prepared COSMOS2020 and JADES DR2 photometry catalogs, which streamlined the workflow for Mephisto. A concerted effort to modernize astronomical tools and data infrastructure, rewriting them in a format that AI agents can readily interact with, will be important for realizing the full potential of agent-based research in astronomy.

This study should not be interpreted as suggesting that all astronomical research can, or even should, be autonomously performed by AI agents. If anything, the difficulties we encountered, even with the relatively simplistic task of SED fitting, are a testament to the ongoing complexity of scientific discovery and the continued importance of human expertise.

\section{Conclusion}\label{sec:con}


In this study, we introduce Mephisto—a multi-agent framework driven by large language models (LLMs) that emulates human-like reasoning to interpret multi-band galaxy observations through spectral energy distribution (SED) modeling. Mephisto integrates with the CIGALE codebase  to iteratively refine physical models against observational data. Its core design includes tree search strategies (depth-first for targeted exploration, breadth-first for coverage) for hypothesis testing, paired with two complementary memory systems: temporal memory (to avoid redundant attempts and prioritize paths within individual analyses) and an external knowledge base (to accumulate validated, domain-specific insights from past sessions). Unlike black-box machine learning approaches in astronomy, Mephisto offers transparent, traceable reasoning that aligns with existing research workflows, serving as a practical research copilot for astronomers. Our key findings are summarized as follows:

\begin{itemize}
\item On normal galaxies, Mephisto matches the performance of exhaustive grid searches (a standard in SED modeling) while using only $\sim1\%$ of the computational resources. Testing on 256 uniformly sampled galaxies from the COSMOS2020 catalog (spanning redshifts 0.1–5.7, stellar masses $10^6–10^{12}\, \mathrm{M}_\odot$, and diverse dust attenuation/star formation rates), its derived physical parameters (e.g., stellar mass, dust attenuation $\mathrm{A}_\mathrm{V}$) show consistency with grid search results.

\item On astrophysical frontiers, Mephisto  analyzes "Little Red Dots" (LRDs)—31 high-redshift galaxies recently discovered by the James Webb Space Telescope (JWST) that postdate most LLMs’ knowledge cutoffs. It independently identifies the two leading hypotheses for LRDs (obscured AGNs with dust-reddened emission or dust-obscured starburst galaxies) and produces results consistent with human expert analyses, demonstrating its ability to reason about novel, unstudied phenomena.

\item Mephisto’s memory and learning systems enhance its efficiency. Ablation studies show temporal memory (tracking past model modifications) leads to a 25\% greater improvement in fit quality (via relative \(\chi^2\)) by the third inference depth compared to memory-disabled runs. Its external knowledge base—built from 72 validated insights (e.g., "adjusting AGN fraction improves mid-infrared fits")—reduces the failure rate (defined as \(\chi^2 > 1.2\times\) baseline) by 11\% when applied to unseen galaxies.

\item Mephisto is scalable for large-scale surveys via cost-effective LLMs. While high-cost GPT-4o serves as the baseline, mid-cost models (e.g., QwQ-32B, at $\sim$1/10 the cost of GPT-4o) achieve 80–90\% success rates (matching the baseline’s performance), and low-cost models (e.g., DeepSeek-R1-Distill-Llama-8B, at $\sim$1/100 the cost) still provide reasonable results for many galaxies. This makes SED fitting for billions of sources (from future missions like Roman, Euclid, or CSST) financially viable.
\end{itemize}

In summary, Mephisto establishes a new paradigm for AI-augmented astronomical research by integrating LLM-powered reasoning with domain-specific scientific research platforms—specifically, the multi-band photometry data and SED modeling tools employed in this work. It not only boosts efficiency but also preserves the transparency important to scientific rigor, empowering astronomers to validate, guide, and build upon its conclusions. 

Consider a Venn diagram. One circle represents the problems that AI agents can currently solve (a circle that will undoubtedly continue to expand). The other circle contains some of the highest-profile and most valuable problems in astronomy. While the overlap between these two circles may remain relatively small at present, this study demonstrates that the intersection is at least non-zero. Mephisto showcases a concrete example of how agent-based research can contribute meaningfully to astronomical discovery. And as performant agents become exponentially cheaper and more capable, this intersection will only continue to grow, paving the way for a future where human researchers and AI collaborators work together to expediting discovery in astronomical research.




\section{Data \& Code Availability}

The data and code supporting this study are publicly available. The primary repository for this work is Zenodo \hyperlink{https://zenodo.org/records/15589045}{https://zenodo.org/records/15589045}; the JADES data were obtained from \hyperlink{https://dx.doi.org/10.17909/8tdj-8n28}{https://dx.doi.org/10.17909/8tdj-8n28}, and the COSMOS2020 catalog from \hyperlink{https://cosmos2020.calet.org}{https://cosmos2020.calet.org}.


\begin{acknowledgments}
ZS and SH gratefully acknowledge financial support from the National Natural Science Foundation of China (Grants Nos. 12273015 and 12433003) and the Space Application System of the China Crewed Space Program. ZC acknowledges support from the National Key R\&D Program of China (grant no. 2023YFA1605600) and Tsinghua University Initiative Scientific Research Program (No. 20223080023). The authors acknowledge Pablo G. P{\'e}rez-Gonz{\'a}lez for the kind support on the original data in \cite{PABLO2024}. The authors acknowledge Guang Yang for his support with the CIGALE software. We extend special thanks to Leyao Wei and Ben Wang for their collaborative efforts in providing scientific demonstrations for the Mephisto project. The authors are particularly grateful to Shuang Zeng for her insightful advice and acknowledge Yunjing Wu, Xiaojing Lin, and Mingyu Li for their valuable discussions.
\end{acknowledgments}

\appendix

\section{Mephisto's Prompt Example}\label{appendix:prompt}



\subsection{Input State}\label{appendix:input_prompt}
We here provide a detailed example of the input state used by Mephisto in its SED fitting process. The input state, formatted in JSON, includes three key components: observational data, the CIGALE model configuration, and the fit quality metrics.

The dataset $d$ encompasses redshift measurements (either from spectroscopic determination or from photometric redshift) and photometric details such as effective wavelengths, bandwidths, fluxes, and signal-to-noise ratios. Due to the current AI limitations in interpreting scientific visuals quantitatively \cite[][]{LILEI2024,YUE2023}, we input data numerically as tuples—a deviation from human researchers' reliance on visual analysis, yet a practical solution for AI agents.

The SED model $m$, constructed in CIGALE, integrates multiple physical components like star formation histories and dust attenuation. The fitting outcome $r$ provides parameter estimates, residuals, computational cost, and fit quality per photometric band, assessed by the agent as good, overestimate, or underestimate. This qualitative method reduces AI numerical hallucinations \cite[][]{YUAN2023,AHN2024}, guiding Mephisto with textual cues for each filter evaluation. It also enables Mephisto to apply domain knowledge to derive more physical changes, such as modifying dust attenuation physical models for suboptimal optical and UV fits.

The model input is then parsed and run by another agent designed to execute CIGALE codes. The results are evaluated by the reasoning prompt, where they are appended as the new state in the subsequent iteration of the tree search.

Below is an example of the input state structure:

\begin{questionbox}[boxcolor2]
\begin{verbatim}
{"data": {
            "redshift": 0.73,
            "photometry": [
                {
                    "name": "hst.wfc.F435W",
                    "wave_eff": 4337,
                    "bandwidth": 940,
                    "band_location_type": "OPTICAL",
                    "band_width_type": "BROAD",
                    "fit_quality": "good",
                    "signal_to_noise": "reliable"
                },
                ...]},     
        "cigale_model": {
            "sfh": {
                "name": "sfhdelayed",
                "params": {
                "tau_main": [100, 500, 1000, 3000, 5000],
                "age_main": [100, 500, 1000, 3000, 5000],
                "tau_burst": [50],
                "age_burst": [20],
                "f_burst": [0.0]}},...},     
        "cigale_best_parameters": {
            "sfh": {
                "name": "sfhdelayed",
                "params": {
                "tau_main": 100,
                "age_main": 500,
                "tau_burst": 50,
                "age_burst": 20,
                "f_burst": 0.0}},...},     
        "cigale_fit_quality": {
        "num_of_good_fit": 5,
        "sum_chi2": 218,
        "grid_size": 450,
        "cigale_message": "CIGALE run smoothly"}}
\end{verbatim}
\end{questionbox}

\subsection{Reasoning Prompt}\label{appendix:reasoning_prompt}
This section presents the reasoning prompt used by Mephisto to guide its decision-making process in SED fitting. The prompt outlines the rules and knowledge base that  Mephisto adheres to when modifying CIGALE models to improve fit quality across different photometric bands. The prompt includes guidelines for module selection, parameter grid specification, and mandatory module requirements. It incorporates two key components that enable Mephisto to emulate human expert reasoning and learning:

\begin{itemize}
    \item \textbf{Temporary Memory:} This feature captures information from the current tree search iteration. It allows Mephisto to adapt its strategies within the ongoing analysis. The prompt instructs Mephisto to consider these temporary memories to ensure diversity and superior fit quality in the updated CIGALE model. The model is updated with information from the current search process, enabling Mephisto to refine its approach based on previous suggestions within the same analysis, avoiding redundant proposals.
    \item \textbf{CIGALE SED Knowledge Base:} This component represents accumulated expertise from previous interactions with data. It enhances Mephisto's ability to improve fit quality for different photometric bands in the CIGALE model. The prompt instructs Mephisto to consider this information when designing the CIGALE model. The agent dynamically filled with relevant expert knowledge about SED fitting, allowing Mephisto to make informed decisions based on established astrophysical principles and learned insights from past analyses. 
\end{itemize}

The integration of temporary memory and the CIGALE SED knowledge base allows Mephisto to generate diverse and reasonable model modifications, combining established principles with both short-term adaptive strategies and long-term learned insights. Importantly, the results of each analysis are not discarded but are instead processed to update the knowledge base. This cycle of learning and integration allows Mephisto to continuously refine its knowledge base, improving its performance over time and across different analyses.

The prompt's output format guides Mephisto to produce structured, diverse, and reasonable CIGALE model modifications. Each proposed modification includes a ``thinking'' step, module name, specific choice, and parameter grid values.

\begin{questionbox}[boxcolor1]

$<$\texttt{CIGALE} Documentation$>$

$<$Introduction to User Input$>$

Your task is to analyze user input and adhere to below rules and knowledges to modify the provided \texttt{CIGALE} models to improve the fit quality of different bands.
You should identify the \texttt{CIGALE} module which should affect the fitting quality the most and carefully modify this module, e.g., use another choice, adjust the parameter grid, and etc.

\begin{itemize}
\item Choice Selection: For each module in the model, select either one choice or none. This decision should be based on the need to optimize the model’s fit for the observational data.
\item Parameter Grid Specification: For the selected choice in each module, define a parameter grid that the model will use to fit the data.
\begin{itemize}
\item For discrete parameters, select grid values from a pre-defined list.
\item For continuous parameters, derive grid values within a specified range to ensure a detailed exploration of parameter space.
\end{itemize}

\item Mandatory Modules: The model configuration must include specific settings for the following modules:
\begin{itemize}
\item sfh (Star Formation History): This module is needed for modeling the rate at which a galaxy forms stars over time.
\item ssp (Simple Stellar Population): This module is important for understanding the collective properties of stars in a galaxy that formed at the same time and with the same metallicity.
\end{itemize}
\item For 'imf' parameter in 'bc03' and 'm2005', it should be 0 or 1, not a list, i.e., "imf": 1 are accepteable, "imf": [1] are foridden
\item  For 'disk\_type' parameter in 'fritz2006' and 'skirtor2016' it should be 0 or 1, not a list, i.e., "disk\_type": 0 are acceptable, "disk\_type": [1] are forbidden
\end{itemize}

\#\#CIGALE SED Knowledge\#\#

To enhance the fit quality for different photometric bands in the CIGALE model, consider the following additional information when designing your CIGALE model.

\%\%KNOWLEDGE\%\%

$\quad$

\#\#Temporary Memory\#\#

Please take into account the following temporary memories to ensure diversity and superior fit quality in the updated CIGALE model:

\%\%MEMORY\%\%

$\quad$

\#\#Expected Output Format\#\#

Your output format should be structured as below list constructed by four diverse and reasonable CIGALE model modifications:

\begin{verbatim}
[
    {
        "thinking": "thinking for CIGALE model modification 0 here"
        "module": "module name for CIGALE model modification 0 here",
        "name": "module choice name for CIGALE model modification 0 here",
        "parameters": [
            "parameter 1": ["parameter 1 grid here"],
            "parameter 2": ["parameter 2 grid here"],
            ...
            "parameter n": ["parameter n grid here"]
        ]
    },
    ...
]
\end{verbatim}
\end{questionbox}

\subsection{Learning Prompt}\label{appnedix:learning_prompt}
This section details the prompt used in the learning process implemented in Mephisto, focusing on how the system acquires, validates, and integrates knowledge to enhance its performance in Spectral Energy Distribution (SED) fitting. The external knowledge base in Mephisto is formatted in natural language, which is important for navigating the complex hypothesis space of SED models, illustrated supplementary ``\texttt{knowledge.json}" file.

During the reasoning process, as described in the previous appendix, Mephisto generates various suggestions for modifying the current SED model to address discrepancies between predictions and empirical data. These refined models are computed using CIGALE and assessed for fit quality improvement. The interaction history is logged and processed by a learning agent, which extracts insights to optimize future SED model designs.

To ensure the quality and relevance of the acquired knowledge, we have implemented a validation and distillation process:

\begin{enumerate}
    \item Knowledge Extraction: The learning agent analyzes the interaction history to synthesize insights for improving CIGALE models. These insights are formulated as universally relevant knowledge that can address inherent design flaws in the models.
    \item Knowledge Validation: A validation agent compiles each piece of knowledge into a logical expression. These expressions are then evaluated using an eval\_state function to determine their applicability to different observation states. This step ensures that the knowledge is consistent with the provided background information and can be applied across various scenarios.
    \item Impact Assessment: The validated knowledge is evaluated based on its impact on reasoning outcomes across multiple sources. This step helps identify knowledge that consistently improves model performance.
    \item Knowledge Filtering: Following the impact assessment, knowledge with negative impact is filtered out, ensuring that only beneficial knowledge is retained.
    \item Knowledge Integration: The remaining beneficial knowledge is integrated into the existing knowledge base through an optimization agent. This agent analyzes discrepancies between new and existing knowledge, synthesizing them into a cohesive whole.
\end{enumerate}

This iterative learning process allows Mephisto to continuously refine its knowledge base, improving its ability to navigate the complex hypothesis space of SED models and generate more good fits over time.

The following prompt illustrates the knowledge extraction component of this learning process:

\begin{questionbox}[boxcolor2]
\#\# Task: Refine Key Insights from Interaction History for CIGALE Model Enhancement

$\quad$

\#\#\# Objective:

$\quad$

Your task is to synthesize insights from the supplied interaction history, which will be instrumental in shaping the development of upcoming CIGALE models. The primary focus should be on improving their overall performance and effectiveness.\\

The insights you derive must be universally relevant and capable of addressing any inherent design flaws in the models.\\

Your analysis is expected to be consistent with the details provided in the interaction history.\\

$\quad$
$<$Examples$>$
$\quad$

\#\#\# Output Format:

$\quad$

List of distilled Knowledge:
$\quad$
\begin{verbatim}
["knowledge 1", "knowledge 2", ...]
\end{verbatim}
\end{questionbox}

The following prompt illustrates the knowledge validation component of this learning process:

\begin{questionbox}[boxcolor2]
Objective: Derive a logical expression from the given knowledge on spectral energy distribution fitting in galaxy physics to determine the applicability of the knowledge to a specified observation state.\\

Background Information:

<Background Information Here>\\

$\quad$

Evaluation Function: Use the eval\_state function to assess observations:

def eval\_state(state, feature, pattern):

$\quad$

    Parameters:

$\quad$

    - state: Identifier for a multi-band observation.

$\quad$

    - feature: Band name or tuple (Band Location Type, Band Width Type). 'ANY' can be used as a wildcard.

$\quad$

    - pattern: Fit quality criterion (good, overestimated, underestimated).

$\quad$

    Returns:

$\quad$

    - True if the state meets the feature with the specified pattern, otherwise False.

$\quad$

Task: Formulate a logical expression using the eval\_state function to identify multi-band observations where the provided knowledge is applicable.
Note: For intricate knowledge, merge conditions using logical operators such as OR (or) and AND (and).
Note: Employ parentheses to structure complex logical expressions as needed.

$\quad$

$<$Examples$>$\\

$\quad$

Output Format:

\begin{verbatim}
{
    "thinking": "detail your thinking here",
    "expression": "expression here"
}
\end{verbatim}
\end{questionbox}

The following prompt illustrates the knowledge integration component of this learning process:

\begin{questionbox}[boxcolor2]
Your task is to maintain a knowledge base used to guide the design of CIGALE SED models. \\

You will receive a knowledge base and a set of new knowledge. \\

You need to distill the new information and integrate it into the existing knowledge base.\\

\#\# Guidelines:

1. If the newly added knowledge does not conflict with any existing knowledge, add it to the knowledge base.

2. If the newly added knowledge conflicts with an existing piece of knowledge, revise the existing knowledge to make them as consistent as possible.

3. If the newly added knowledge is the same as an existing piece of knowledge, organize and merge them.

4. Make the distilled knowledge consise and informative.\\

\#\# Input Format:

\begin{verbatim}
{
    "BaseKnowledge": [
        "a list of base knowledge here"
    ],
    "KnowledgeGradient": [
        "a list of new knowledge here"
    ]
}
\end{verbatim}

\#\# Output Format:

\begin{verbatim}
{
    "thinking": "Describing the difference 
    between BaseKnowledge and KnowledgeGradient first, 
    and analyze how to combine the two sets of knowledge.",
    "UpdatedKnowledgeBase": [
        "a list of updated knowledge "
    ]
}
\end{verbatim}
\end{questionbox}

\section{Pseudocode for Mephisto's Reasoning Process}\label{appendix:code}

To clarify the reasoning process, we provide detailed pseudocode for Mephisto’s adaptive search strategy, which selects between depth-first search (DFS) and breadth-first search (BFS) to explore the SED model space.

Starting from an initial state, one can select either Breadth-First Search (BFS) or Depth-First Search (DFS) based on scientific objectives and available computational resources. Mephisto then explores the SED model space by generating hypotheses, analyzing discrepancies between model predictions and observations and updating memory, guided by the selected strategy. This process continues until a predefined termination condition is met (e.g., maximum number of explored states). Below are the key details of DFS and BFS:

\begin{enumerate}
\item \textbf{Depth-First Search (DFS)}: This strategy focuses on recursive exploration by expanding the node (defined by a predefined evaluation function). By deeply exploring high-potential modification pathways, DFS uses heuristic guidance to efficiently traverse solution spaces with strong directional biases, pursuing local optima until performance gains diminish \cite[][]{YAO2023}.

\item \textbf{Breadth-First Search (BFS)}: This strategy emphasizes systematic, level-wise exploration, evaluating all adjacent states at the current level before moving to subsequent layers. Using a first-in-first-out queue, BFS ensures coverage of the solution space, exploring diverse modification directions to reduce the risk of missing non-intuitive but viable solutions—critical for domains with high uncertainty or sparse prior knowledge \cite[][]{YAO2023}.
\end{enumerate}

In the following, we have:

\begin{itemize}
    \item $N_{\text{max}}$: Maximum number of total states to explore
    \item $D_{\text{max}}$: Maximum depth of the search tree
    \item $s_0$: Initial state (root of the search tree)
    \item $\mathcal{K}$: Knowledge base
    \item $\mathcal{M}$: Memory of explored states
    \item $R$: Reasoning agent that generates new states
    \item $E$: Evaluation function that assesses new states
\end{itemize}

The pseudocode for the two search strategies (DFS and BFS) is provided in Algorithms 1 and 2.

\begin{algobox}
    \begin{tabular}{@{}l}
        \textbf{Algorithm 1: Depth-First Search Reasoning Process} \\
        \hline
    \end{tabular}
    \begin{enumerate}
        \setlength{\itemindent}{1em}
        \setlength{\itemsep}{0.3em}

        \item \kw{Input}: $s_0$, $\mathcal{K}$, $\mathcal{M}=\{\}$, $R$, $E$, $D_{\text{max}}$, $N_{\text{max}}$
        
        \item \kw{Initialize}: stack $S \gets [s_0]$ \com{DFS stack}
        \item $\text{total\_states} \gets 1$ \com{Explored states count}
        
        \item \kw{While} $S \neq []$ and $\text{total\_states} < N_{\text{max}}$:
            \begin{enumerate}
                \item $s \gets S.\text{pop()}$ \com{Get latest state}
                \item $s'_1,\ldots,s'_n \gets R(s, \mathcal{K}, \mathcal{M})$ \com{Generate new states}
                \item $s''_1,\ldots,s''_n \gets E(s'_1,\ldots,s'_n)$ \com{Evaluate states}
                \item \kw{For} each $s''$ in $s''_1,\ldots,s''_n$:
                    \begin{enumerate}
                        \item $s.\text{add\_child}(s'')$ \com{Add to tree}
                        \item \kw{If} $s''.\text{depth} < D_{\text{max}}$:
                            \begin{enumerate}
                                \item $S.\text{push}(s'')$ \com{Add to stack}
                                \item $\text{total\_states} \gets \text{total\_states} + 1$
                            \end{enumerate}
                    \end{enumerate}
            \end{enumerate}
        
        \item $\mathcal{K} \gets \mathcal{K}.\text{update}(\mathcal{M})$ \com{Update knowledge base}
        \item \kw{Return} $\text{summarize}(s_0)$ \com{Summarize search tree}
    \end{enumerate}
\end{algobox}

\newpage

\begin{algobox}
    \begin{tabular}{@{}l}
        \textbf{Algorithm II: Breadth-First Search Reasoning Process}\label{alg:bfs} \\
        \hline
    \end{tabular}
    \begin{enumerate}
        \setlength{\itemindent}{1em}
        \setlength{\itemsep}{0.3em}

        \item \kw{Input}: $s_0$, $\mathcal{K}$, $\mathcal{M}=\{\}$, $R$, $E$, $D_{\text{max}}$, $N_{\text{max}}$
        
        \item \kw{Initialize}: queue $Q \gets [s_0]$ \com{BFS queue}
        \item $\text{total\_states} \gets 1$ \com{Explored states count}
        
        \item \kw{While} $Q \neq []$ and $\text{total\_states} < N_{\text{max}}$:
            \begin{enumerate}
                \item $s \gets Q.\text{dequeue}()$ \com{Get earliest state}
                \item $s'_1,\ldots,s'_n \gets R(s, \mathcal{K}, \mathcal{M})$ \com{Generate new states}
                \item $s''_1,\ldots,s''_n \gets E(s'_1,\ldots,s'_n)$ \com{Evaluate states}
                \item \kw{For} each $s''$ in $s''_1,\ldots,s''_n$:
                    \begin{enumerate}
                        \item $s.\text{add\_child}(s'')$ \com{Add to tree}
                        \item \kw{If} $s''.\text{depth} < D_{\text{max}}$:
                            \begin{enumerate}
                                \item $Q.\text{enqueue}(s'')$ \com{Add to queue}
                                \item $\text{total\_states} \gets \text{total\_states} + 1$
                            \end{enumerate}
                    \end{enumerate}
                \item $\mathcal{M}.\text{update}(s, s''_1,\ldots,s''_n)$ \com{Update memory}
            \end{enumerate}
        
        \item $\mathcal{K} \gets \mathcal{K}.\text{update}(\mathcal{M})$ \com{Update knowledge base}
        \item \kw{Return} $\text{summarize}(s_0)$ \com{Summarize search tree}
    \end{enumerate}
\end{algobox}

\section{Dataset Details}\label{appnedix:full_data}
\subsection{Dataset Construction of COSMOS2020 Catalog}\label{appendix:cosmos_data}

To evaluate Mephisto's capabilities in spectral energy distribution (SED) modeling, we utilized the COSMOS2020 catalog \cite[][]{WEAVER2022}, which provides a compilation of photometric observations across a wide range of wavelengths. Our initial dataset comprised 723,897 photometric observations. To enhance the quality and reliability of our analysis, we cross-matched the COSMOS2020 catalog with several spectroscopic surveys, including PRIMUS \cite[][]{PRIMUS2013}, 3D-HST \cite[][]{BRAMMER2012}, SDSS DR16 \cite[][]{SDSSDR162020}, C3R2 \cite[][]{C3R22019}, DEIMOS \cite[][]{DEIMOS2018}, MOSDEF \cite[][]{KRIEK2015}, and LEGA-C \cite[][]{VANDERWEL2021}. This cross-matching process resulted in a refined sample of 38,547 sources with high-quality spectroscopic redshifts and ancillary data.

\begin{figure*}[htbp]
    \centering
    \subfloat[Dust Attenuation, $\mathrm{A}_\mathrm{V}$ \textit{v.s.} Stellar Mass, log $\mathrm{M}_{*}/\mathrm{M}_{\odot}$ \label{fig:cosmos_demo_aa}]{
        \includegraphics[width=0.45\linewidth]{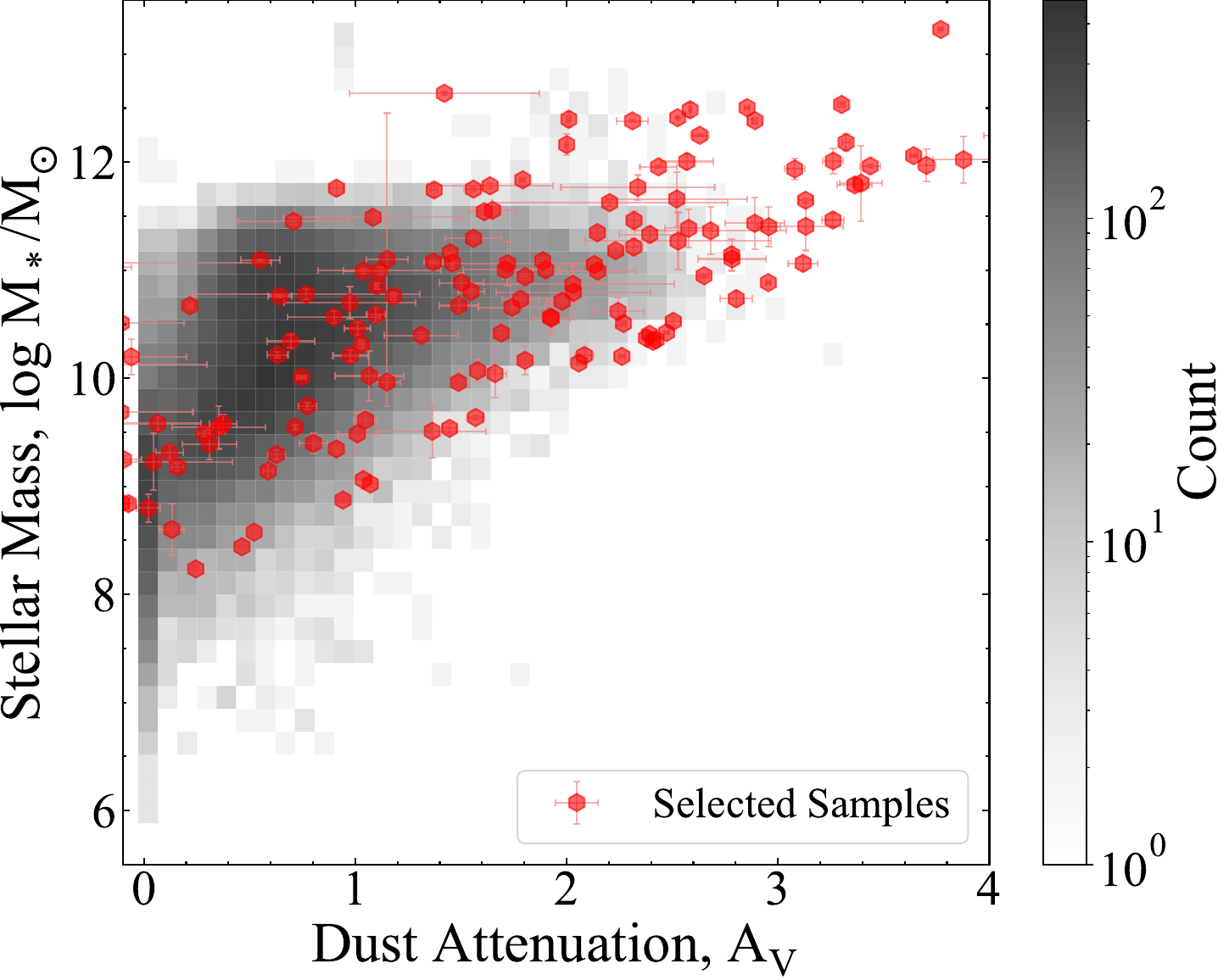}
    }
    \hfill
    \subfloat[Star Formation Rate, SFR \textit{v.s.} Stellar Mass, log $\mathrm{M}_{*}$ \label{fig:cosmos_demo_bb}]{
        \includegraphics[width=0.45\linewidth]{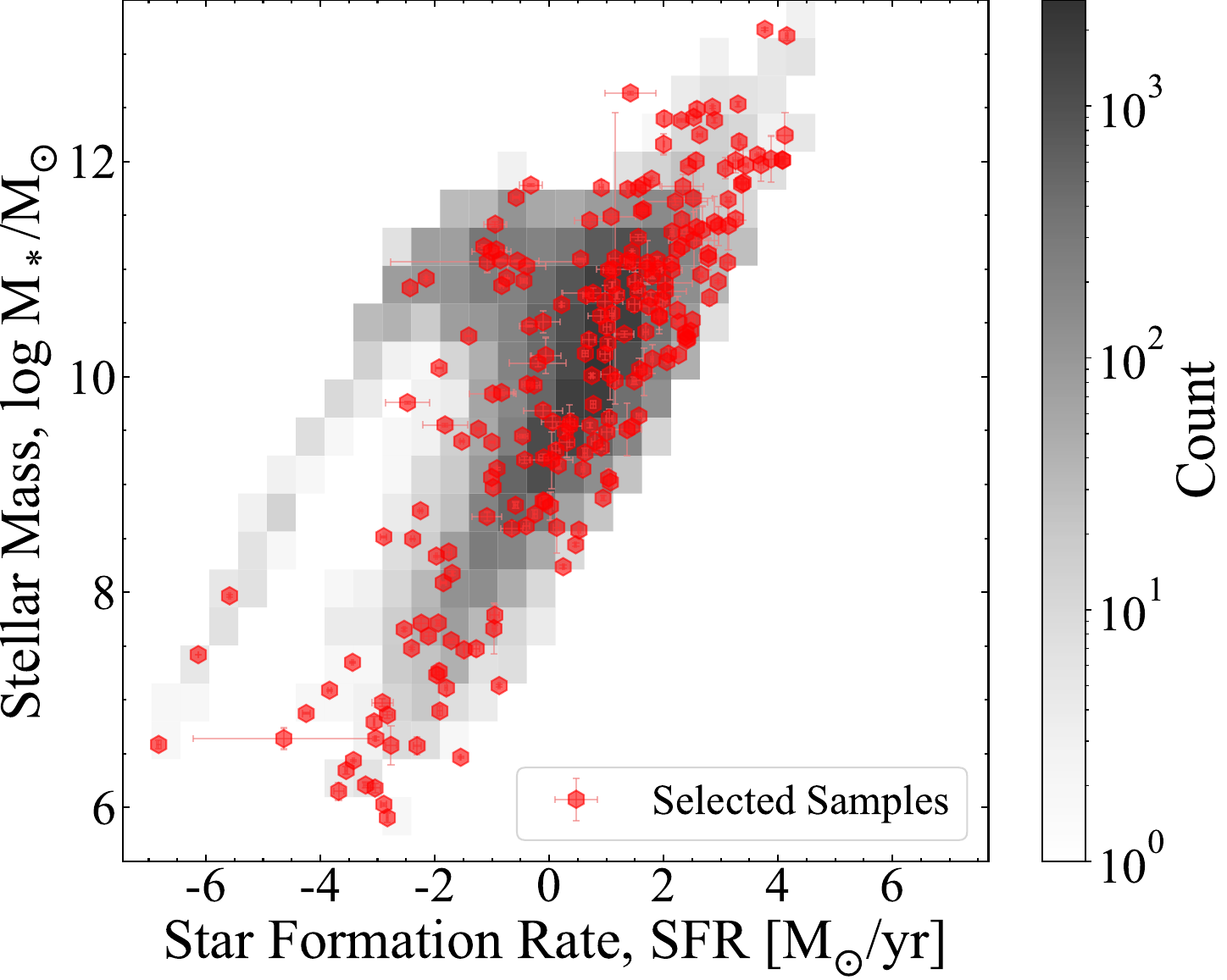}
    }
    \caption{Distribution of physical properties in the COSMOS2020 catalog and our selected sample. The two-dimensional grayscale histograms represent the density distribution of galaxies in the original COSMOS2020 catalog, with darker regions indicating higher concentrations of objects. Red points denote our selected sample of 256 galaxies. \textbf{(a)} Distribution in the parameter space of dust attenuation ($A_V$) versus stellar mass ($\log M_*/M_\odot$). \textbf{(b)} Distribution in the parameter space of star formation rate (SFR, in $M_\odot/\mathrm{yr}$) versus stellar mass ($\log M_*/M_\odot$). The selected sample exhibits a broad and uniform distribution across the full range of physical properties, spanning approximately four orders of magnitude in stellar mass ($10^6$--$10^{13} M_\odot$), the complete observed range of dust attenuation values ($A_V \approx 0$--$4$), and more than ten orders of magnitude in star formation rate ($10^{-6}$--$10^{6} M_\odot/\mathrm{yr}$). This deliberate sampling strategy ensures coverage of the parameter space, including regions with low galaxy number densities that would be underrepresented in a random selection. The uniformity of our sample across these key physical parameters is important for evaluating Mephisto's performance across diverse galaxy populations and environmental conditions.}
    \label{fig:cosmos_distribution}
\end{figure*}

To ensure a diverse and representative subset for our experiments, we employed a neural spline flow \cite[][]{DURKAN2019} for density estimation within a four-dimensional parameter space defined by redshift, stellar mass, V-band dust attenuation, and star formation rate (SFR). By estimating the density in this parameter space, we were able to assign weights to each galaxy inversely proportional to the local density, thereby promoting the selection of galaxies from underrepresented regions and ensuring a uniform coverage across the entire parameter space as shown in Figure~\ref{fig:cosmos_distribution}.

\subsection{CIGALE Initial Model}\label{appendix:base_model}
Mephisto commences with a minimalistic spectral energy SED model and systematically explores the SED hypothesis space through iterative incorporation of additional modules via empirical refinement. A constrained initial model is necessary to this process, which we explicitly delineate here. Our baseline SED model incorporates only variable star-formation history and dust attenuation parameters, while deliberately fixing specific parameters such as the extinction law and UV bump properties. This parsimonious framework establishes a foundational physical interpretation of the observational data. Subsequently, Mephisto iteratively revises its internal assumptions through a structured cycle of hypothesis generation and evaluation, addressing discrepancies between model predictions and empirical SEDs through methodical parameter adjustments.

\begin{table}[ht]
\centering
\begin{tabular}{c|c|c}
\hline
\hline
\textbf{Module} & \textbf{Parameter} & \textbf{Value} \\
\hline
\texttt{sfhdelayed} & tau\_main (10$^{6}$ years) & 100, 500, 3000, 5000, 7000, 8000, 9000, 13000 \\
 & age\_main (10$^{6}$ years) & 100, 500, 3000, 5000, 7000, 8000, 9000, 13000 \\
 & tau\_burst (10$^{6}$ years) & 50 \\
 & age\_burst (10$^{6}$ years) & 5, 50, 75 \\
 & f\_burst & 0.001, 0.01, 0.1 \\
\hline
\texttt{bc03} & imf & 1 (Chabrier) \\
 & metallicity & 0.02 \\
\hline
\texttt{nebular} & logU & -2.0 \\
 & f\_esc & 0.0 \\
 & f\_dust & 0.0 \\
 & lines\_width (km s$^{-1}$) & 300 \\
\hline
\texttt{dustatt\_modified\_starburst} & E\_BV\_nebular (mag) & 0.005, 0.02, 0.05, 0.1, 0.5, 0.7\\
 & E\_BV\_factor & 0.50 \\
 & uv\_bump\_wavelength (nm) & 217.5 \\
 & uv\_bump\_width (nm) & 35.0 \\
 & uv\_bump\_amplitude & 0.0 \\
 & powerlaw\_slope & 0.0 \\
 & Ext\_law\_emission\_lines & 1 (Milky Way) \\
 & Rv & 3.1 \\
 & filters & FUV, V\_B90 \\
\hline
\texttt{dl2014} & qpah & 2.5 \\
& umin & 1.0 \\
& alpha & 2.0 \\
& gamma & 0.1 \\
\hline
\texttt{redshifting} & redshift &  \\
\hline
\end{tabular}
\caption{Initial Spectral Eenergy Distribution Model}
\label{tab:base}
\end{table}

\subsection{CIGALE Baseline Model}\label{appendix:baseline}

To establish a comparative baseline, we employed the Code Investigating GALaxy Emission (CIGALE) \cite[][]{CIGALE2019} for spectral energy distribution modeling of the selected 256 galaxies. CIGALE represents a widely adopted astrophysical tool for modeling multi-wavelength galaxy emission by fitting photometric data to theoretical SED templates. Given the substantial diversity of galactic properties within our sample, we configured CIGALE with a comprehensive and finely sampled parameter grid to capture the complex variations in galaxy characteristics.

Our baseline model was constructed by systematically curating the SED grid based on the framework established by Brown et al. (2014) \cite[][]{BROWN2014}, originally designed for modeling 129 local galaxies with heterogeneous properties. To adapt this grid for our more expansive and diverse sample spanning redshifts from 0.1 to 5.7, we implemented a fine-grained parameterization for the main stellar population age and AGN emission fraction within the dust emission model. This methodical refinement enables the baseline model to account for evolutionary stages and AGN contributions across a broader spectrum of galaxy morphologies and cosmic epochs.

The resultant baseline SED grid, as detailed in Table~\ref{tab:baseline}, encompasses approximately $3.63 \times 10^8$ models, incorporating 10 free parameters including stellar age, metallicity, dust attenuation, star formation history, and AGN fraction. This grid ensures adequate model space density to provide good fits for the majority of galaxies in our sample. By utilizing this defined baseline, we can execute a direct quantitative comparison between results obtained via Mephisto and those derived from traditional exhaustive grid search methodologies commonly implemented in large-scale sky surveys for population-level SED analysis.

\begin{table}[ht]
\centering
\begin{tabular}{c|c >{\centering\arraybackslash}p{8cm}}
\hline
\hline
\textbf{Module} & \textbf{Parameter} & \textbf{Value} \\
\hline
\texttt{sfhdelayed} & tau\_main (10$^{6}$ years) & 1, 500, 1000, 2000, 3000, 4000, 5000, 6000, 7000, 8000, 9000, 10000, 11000, 12000, 13000 \\
 & age\_main (10$^{6}$ years) & 1000, 2000, 3000, 4000, 5000, 6000, 7000, 8000, 9000, 10000, 11000, 12000, 13000 \\
 & tau\_burst (10$^{6}$ years) & 13000 \\
 & age\_burst (10$^{6}$ years) & 5, 10, 25, 50, 100, 200, 350, 500, 750, 900 \\
 & f\_burst & 0, 0.0001, 0.0005, 0.001, 0.005, 0.01, 0.05, 0.1, 0.25 \\
\hline
\texttt{bc03} & imf & 1 (Chabrier) \\
 & metallicity & 0.02 \\
\hline
\texttt{nebular} & logU & -3.0 \\
 & f\_esc & 0.0 \\
 & f\_dust & 0.0 \\
 & lines\_width (km s$^{-1}$) & 300 \\
\hline
\texttt{dustatt\_modified\_starburst} & E\_BV\_nebular (mag) &0.005, 0.01, 0.025, 0.05, 0.075, 0.10, 0.15, 0.20, 0.25, 0.30, 0.35, 0.40, 0.45, 0.50, 0.55, 0.60\\
 & E\_BV\_factor & 0.25, 0.50, 0.75 \\
 & uv\_bump\_wavelength (nm) & 217.5 \\
 & uv\_bump\_width (nm) & 35.0 \\
 & uv\_bump\_amplitude & 0.0, 1.5, 3.0 (Milky Way) \\
 & powerlaw\_slope & -0.5, -0.4, -0.3, -0.2, -0.1, 0.0 \\
 & Ext\_law\_emission\_lines & 1 (Milky Way) \\
 & Rv & 3.1 \\
 & filters & FUV, V\_B90 \\
\hline
\texttt{dale2014} & alpha & 0.5, 1.0, 1.5, 2.0, 3.5, 3.0, 3.5, 4.0 \\
& fracAGN & 0., 0.1, 0.99\\
\hline
\texttt{redshifting} & redshift &  \\
\hline
\end{tabular}
\caption{Baseline Spectral Eenergy Distribution Model}
\label{tab:baseline}
\end{table}

\section{Extended Scientific Applications}

Beyond the ``Little Red Dots'' sample presented in the main text, we demonstrate two additional scientific applications of Mephisto at the astrophysical research frontier, each addressing distinct scientific questions. At the time of writing, Mephisto has been shared with select trusted research groups to evaluate its performance and ensure its successful application to various out-of-domain questions.

\subsection{Application to an Isolated Ultra-Diffuse Galaxy Candidate: ``Zangetsu"}

\begin{figure*}
    \centering
    \subfloat[SED fitting results for Zangetsu]{
        \includegraphics[width=0.45\linewidth]{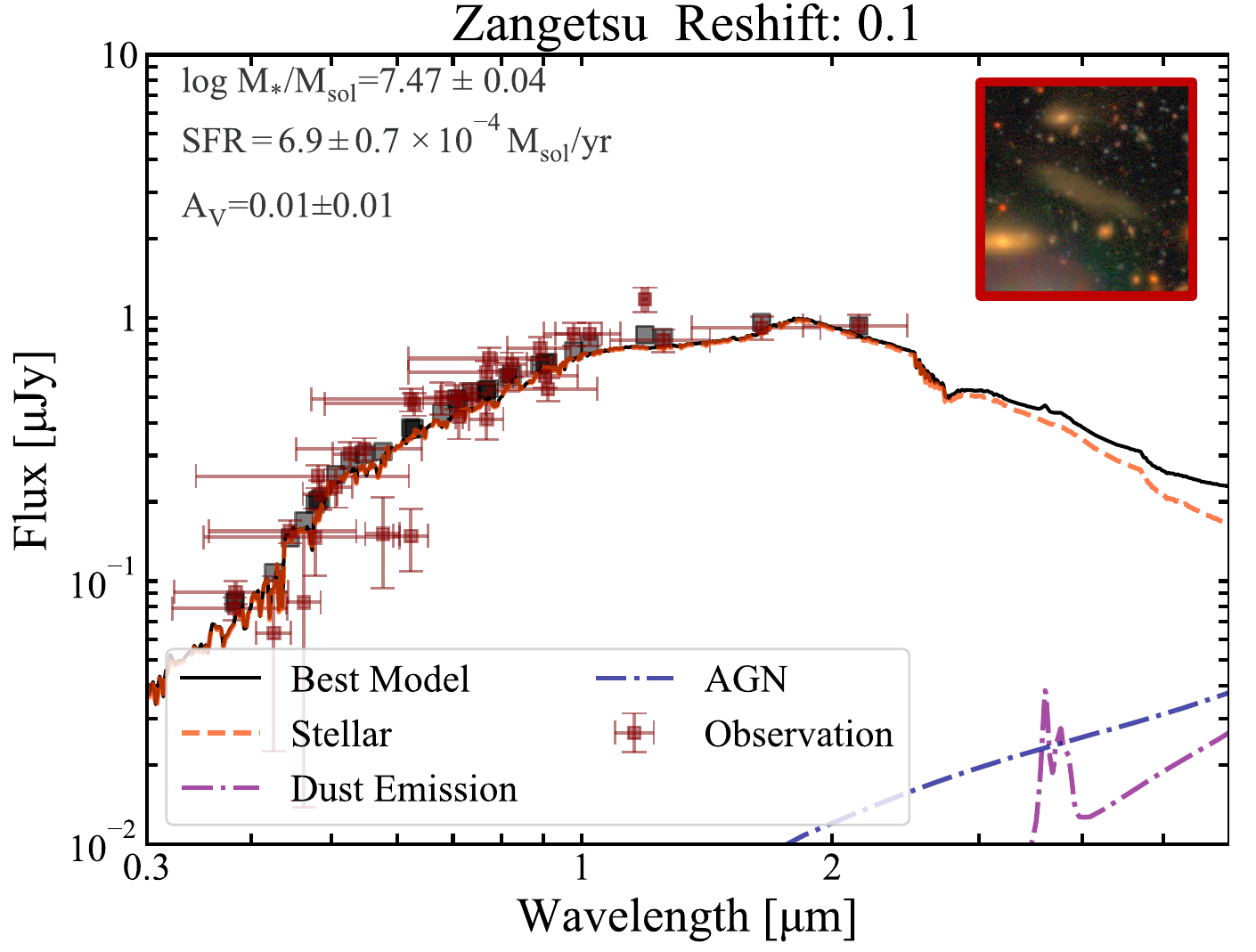}
        \label{fig:zangetsu_sed}
    }
    \hfill
    \subfloat[Stellar mass $\log\,\mathrm{M}_{*}$ \textit{v.s.} star-formation rate $\log\,\mathrm{SFR}$ for various SED models evaluated by Mephisto]{
        \includegraphics[width=0.45\linewidth]{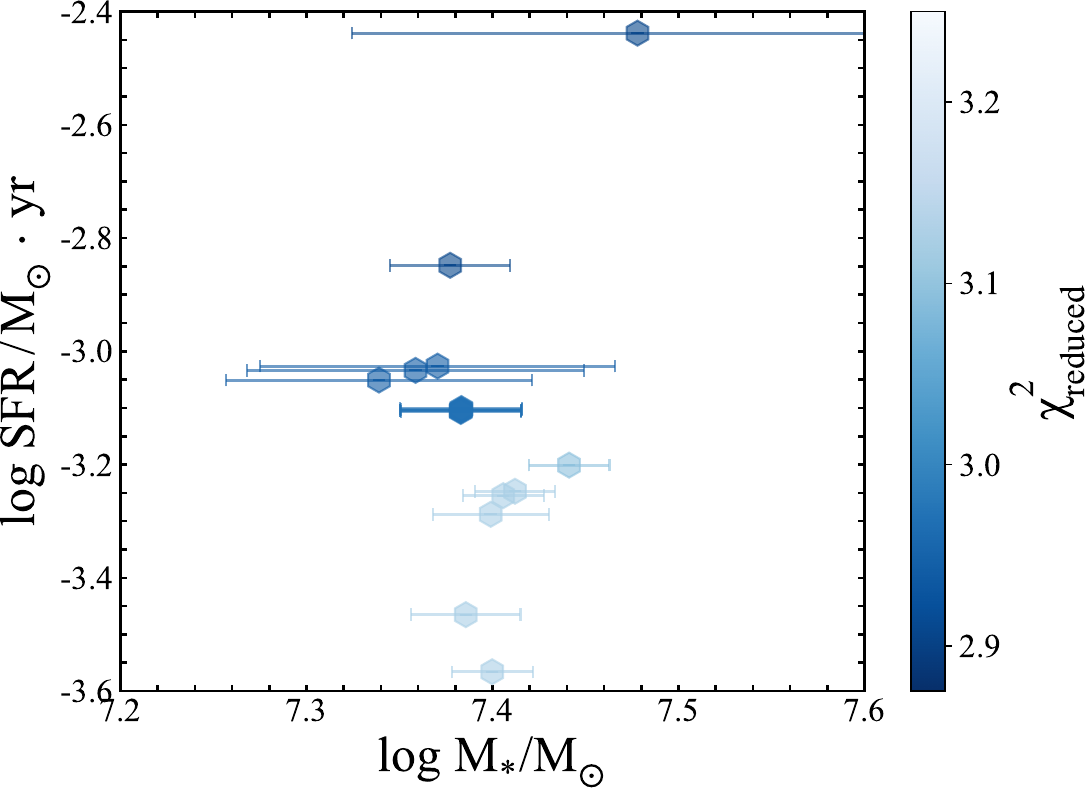}
        \label{fig:zangetsu_solution}
    }
    \caption{Analysis of Zangetsu's spectral energy distribution using Mephisto. \textbf{(a)} Observed multi-wavelength photometry (brown squares with error bars) and the corresponding best-fit SED model (solid black line). Component contributions are delineated as follows: stellar emission (dashed red line), dust emission (dot-dashed purple line), and AGN contribution (dot-dashed blue line). The inset image shows the HSC-SSP optical observation of Zangetsu, revealing its elongated, low surface brightness morphology. Key derived physical parameters are displayed in the upper left: stellar mass ($\log M_*/M_\odot = 7.47 \pm 0.04$), star formation rate ($\mathrm{SFR} = 6.9 \pm 0.7 \times 10^{-4}\ M_\odot/\mathrm{yr}$), and dust attenuation ($A_V = 0.01 \pm 0.01$). \textbf{(b)} Posterior SED solutions sampled by Mephisto, projected onto the stellar mass–star-formation rate ($\log M_*$–$\log \mathrm{SFR}$) parameter space. Points are color-coded by their reduced $\chi^2$ values, with darker blue indicating better fits. All solutions consistently indicate Zangetsu's quiescent nature, with specific SFR values ($\mathrm{SFR}/M_*$) approximately three orders of magnitude below the star-forming main sequence at similar redshift.}
    
    \label{fig:zangetsu}
\end{figure*}

We employ Mephisto to analyze ``Zangetsu," an isolated ultra-diffuse galaxy (UDG) candidate in the COSMOS field recently reported by Wei et al. 2025 \cite[][]{WEI2025}. This peculiar object, observed through the Hyper Suprime-Cam Subaru Strategic Program (HSC-SSP), exhibits the following key characteristics as shown in Figure~\ref{fig:zangetsu_sed}: (1) A flat surface brightness profile (Sérsic index $n \sim 0.4$); (2) Very low central surface brightness ($i \sim 25.53$ mag arcsec$^{-2}$); (3) Unusual elongation ($b/a \sim 0.25$); (4) Asymmetric morphology with distortion features.
Its red optical color ($g - i \sim 0.8$ mag) further suggests a quiescent stellar population. 

To validate this hypothesis, we perform spectral energy distribution (SED) modeling using Mephisto as shown in Figure~\ref{fig:zangetsu}. We utilize photometric data from the COSMOS2020 catalog \cite[][]{WEAVER2022}, wherein Mephisto evaluates multiple SED models as illustrated in Figure~\ref{fig:zangetsu_solution}. Mephisto determines the stellar mass to be $\sim2.5\times 10^{7}\,\mathrm{M}_{\odot}$. The star formation rate analysis confirms the galaxy's predominantly quiescent nature (hence the name ``Zangetsu''), though values vary between different SED models.

\subsection{Application to Composite Photometry of 23 Type-II Quasars}

We employ Mephisto to analyze the composite photometry of 23 Type-II quasars presented in Wang et al. 2025 \cite[][]{BENWANG2025}. The photometric data in this study were collected from multiple imaging surveys, including PS1, DESI Legacy Survey, UKIDSS, Spitzer, and WISE, encompassing a broad wavelength range from ultraviolet (UV) to infrared (IR). 

To generate the composite photometry, we initially obtained spectral energy distribution (SED) model fits for each of the 23 Type-II quasars in their sample. The photometric measurements for individual targets were subsequently scaled using the luminosity at rest-frame 30 $\mu$m—a wavelength devoid of emission or absorption lines—and all targets were normalized to a fixed luminosity value of $10^{46.33}$ erg/s at this reference wavelength. Following this normalization procedure, the individual photometric measurements were aggregated into 20 logarithmically spaced wavelength bins, and the mean luminosity within each bin was calculated. 

This methodical approach facilitated the construction of composite photometry, represented as a series of data points spanning the entire wavelength range, providing an unified perspective of quasar emission patterns. The resultant composite photometry reveals the collective luminosity distribution of the sample, offering insights into the characteristic features of these heavily obscured active galactic nuclei.

\begin{figure*}
    \gridline{
    \fig{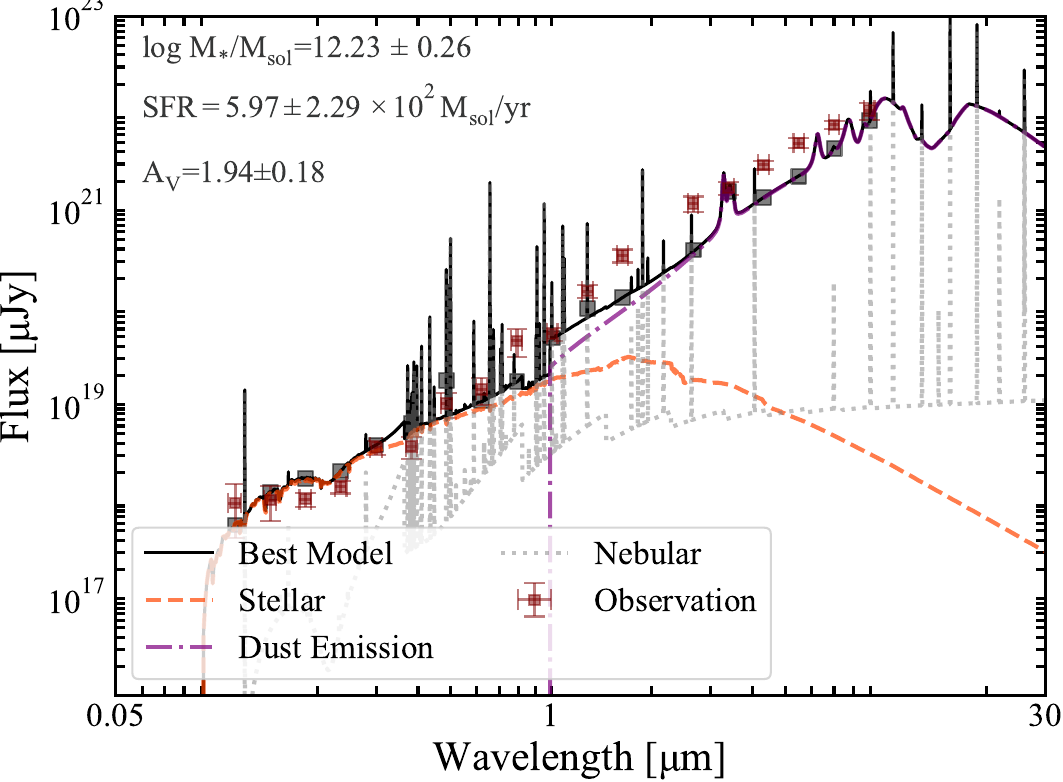}{0.45\textwidth}{SED Solution with only stellar contribution]}
    \fig{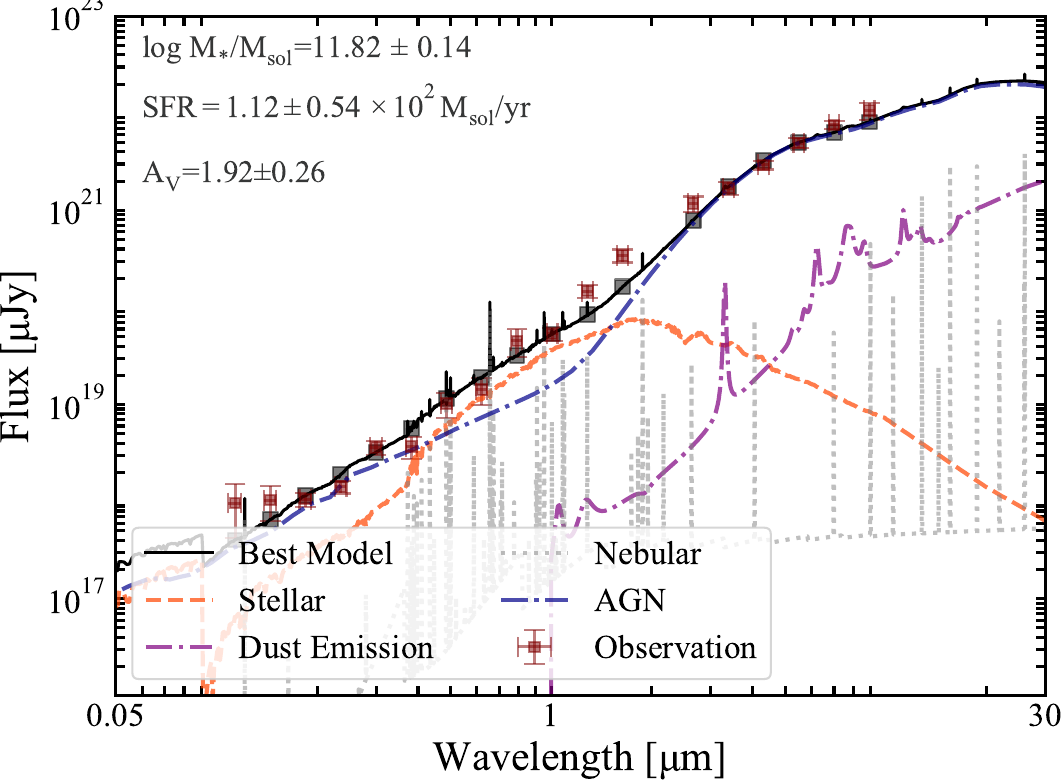}{0.45\textwidth}{SED Solution with mixed contribution from both stellar and active galactic nuclei}
    }
    \centering
    \caption{Alternative spectral energy distribution models for the composite Type-II quasar sample. \textbf{(a)} SED solution derived using Mephisto, assuming exclusively stellar emission contributes to the observed photometry. The best-fit model (solid black line) consists of stellar emission (dashed red line) and dust emission (dot-dashed purple line) components, with nebular emission shown as dotted gray line. Key derived physical parameters are displayed in the upper left: stellar mass ($\log M_*/M_\odot = 12.23 \pm 0.26$), star formation rate ($\mathrm{SFR} = 5.97 \pm 2.29 \times 10^2\ M_\odot/\mathrm{yr}$), and dust attenuation ($A_V = 1.94 \pm 0.18$). \textbf{(b)} SED solution integrating contributions from both stellar populations and active galactic nuclei, with the AGN component (dot-dashed blue line) modeled as a reddened accretion disk. This model yields slightly different physical parameters: stellar mass ($\log M_*/M_\odot = 11.82 \pm 0.14$), star formation rate ($\mathrm{SFR} = 1.12 \pm 0.54 \times 10^2\ M_\odot/\mathrm{yr}$), and dust attenuation ($A_V = 1.92 \pm 0.26$). Both solutions are superimposed on the composite photometry (brown squares with error bars) constructed from multiple imaging surveys spanning UV to IR wavelengths (0.05-30 $\mu$m). The logarithmic wavelength scale highlights spectral features across four orders of magnitude.}
    \label{fig:composite_sed}
\end{figure*}

As illustrated in Figure~\ref{fig:composite_sed}, the composite SED exhibits pronounced emission in the ultraviolet and optical bands. We propose three potential physical mechanisms to interpret this phenomenon: (1) reddened accretion disk emission; (2) scattered light; (3) stellar emission. While scattered light phenomena are not incorporated in CIGALE model libraries, application of Mephisto to this composite SED derived two plausible physical scenarios involving reddened accretion disk and stellar emission contributions, as demonstrated in Figure~\ref{fig:composite_sed}. The star formation rate derived by Mephisto aligns with the stellar emission scenario conjecture independently established through the AGNfitter SED modeling tool by human researchers \cite[][]{BENWANG2025}.

\section{Examples for knowledge distilled by Mephisto}\label{appendix:knowledge}
Here we present examples from Mephisto’s knowledge base, derived via its exploration process through spectral energy distribution (SED) modeling of 208 galaxies in the COSMOS2020 training set. These refined insights focus on fine-grained adjustments to CIGALE physical models, capturing empirical relationships between observed flux discrepancies and SED model selections. Notable entries include:

\begin{enumerate}
    \item If the model struggles to match observations in near- to mid-infrared bands, try slightly increasing the \texttt{fracAGN} parameter (the fraction of emission from the active galactic nucleus). The AGN component influences the mid-infrared spectrum, so this tweak may improve the fit.
    \item If your model includes the \texttt{nebular} module (for nebular emission) but it doesn’t improve the fit—especially when only broad-band photometry is used (where nebular emission is typically not dominant)—consider removing it.
    \item If the model significantly underestimates flux in near- to short-infrared bands, expand the \texttt{separation\_age} parameter grid in the \texttt{ssp} (single stellar population) module. Older stellar populations often contribute more to these bands, so a wider age range may better capture their effects.
    \item If the model badly underpredicts optical bands, adjust the dust attenuation parameters—specifically \texttt{Av\_ISM} (dust attenuation in the interstellar medium) in the \texttt{dustatt\_modified\_CF00} module. The interstellar medium’s dust may block more or less light than the model initially assumes, so tweaking this parameter could help.
    \item For poor fits in near- to mid-infrared bands, try switching to the \texttt{m2005} (single stellar population) module and broaden its \texttt{separation\_age} and \texttt{metallicity} parameter ranges. This may yield a better match to the observed spectrum.
    \item If infrared bands are poorly fitted, adjust \texttt{dustem} parameters like \texttt{qpah} (PAH mass fraction) and \texttt{umin} (minimum radiation field strength). These parameters directly control dust emission characteristics, so expanding their ranges or testing new values could enhance the model’s performance.
    \item If the model strongly underpredicts mid-infrared bands, adjust the \texttt{EBV} parameter (extinction) in the \texttt{agn} module. Expanding the ranges of both \texttt{EBV} and \texttt{fracAGN} in this module may better account for AGN-related extinction effects and improve the fit.
    \item If the model overestimates flux in optical and near-infrared (NIR) bands while underestimating it in short-wave infrared (SWIR) and mid-wave infrared (MWIR) bands, consider narrowing the parameter grid for \texttt{tau\_main} (star formation time-scale) and \texttt{age\_main} (stellar population age) in the \texttt{sfh} (star formation history) module to prioritize older stellar populations.
\end{enumerate}

This knowledge, expressed in natural language, can be transferred between various large language model backbones. It can also be refined or corrected by human researchers and may even guide the design of future SED models.

\bibliography{main}{}
\bibliographystyle{aasjournalv7}



\end{document}